\documentclass[12pt]{article}
\usepackage{ascmac}
\usepackage{ulem}
\usepackage{latexsym}
\usepackage{graphicx}
\usepackage{amsmath}
\usepackage[top=3.5cm,bottom=3.5cm,left=2cm,right=2cm]{geometry}
\usepackage{pifont}
\usepackage{cancel}             
\usepackage{amsmath,amssymb}

\newcommand{\h}{\hspace}
\newcommand{\be}{\begin{equation}}
\newcommand{\e}{\end{equation}}

\begin{document}

\title{
\vbox{ 
\baselineskip 14pt
\hfill \hbox{\normalsize KUNS-2540 
}} \vskip 1cm
\bf \Large Criticality and Inflation of the Gauged B-L Model\vskip 0.5cm
}
\author{
Kiyoharu~Kawana\thanks{E-mail: \tt kiyokawa@gauge.scphys.kyoto-u.ac.jp}
\bigskip\\
\it \normalsize
 Department of Physics, Kyoto University, Kyoto 606-8502, Japan\\
\smallskip
}
\date{\today}

\maketitle

\abstract{\normalsize
We consider the multiple point principle (MPP) and the inflation of the gauged B-L extension of the Standard Model (SM) with a classical conformality. We examine whether the scalar couplings and their beta functions can become simultaneously zero at $\Lambda_{\text{MPP}}:=10^{17}$ GeV by using the two-loop renormalization group equations (RGEs). We find that we can actually realize such a situation and that the parameters of the model are uniquely determined by the MPP. However, as discussed in \cite{Iso:2012jn}, if we want to realize the electroweak symmetry breaking by the radiative B-L symmetry breaking, the self coupling $\lambda_{\Psi}$ of a newly introduced SM singlet complex scalar $\Psi$ must have a non-zero value at $\Lambda_{\text{MPP}}$, which means the breaking of the MPP. We find that the ${\cal{O}}(100)$GeV electroweak symmetry breaking can be achieved even if this breaking is very small; $\lambda_{\Psi}(\Lambda_{\text{MPP}})\leq10^{-10}$. Within this situation, the mass of the B-L gauge boson is predicted to be
\begin{equation} M_{B-L}=2\sqrt{2}\times\sqrt{\frac{\lambda(v_{h})}{0.10}}\times v_{h}\simeq 696\h{1mm}\text{GeV},\nonumber\end{equation}
where $\lambda$ is the Higgs self coupling and $v_{h}$ is the Higgs expectation value. This is a remarkable prediction of the (slightly broken) MPP. Furthermore, such a small $\lambda_{\Psi}$ opens a new possibility: $\Psi$ plays a roll of the inflaton \cite{Okada:2011en}. Another purpose of this paper is to investigate the $\lambda_{\Psi}\Psi^{4}$ inflation scenario with the non-minimal gravitational coupling $\xi\Psi^{2} {\cal{R}}$ based on the two-loop RGEs.

\newpage

\section{Introduction}
The discovery of the Higgs like particle and its mass \cite{Aad:2012tfa,Chatrchyan:2012ufa} is very meaningful for the Standard Model (SM). The experimental value of the Higgs mass suggests that the Higgs potential can be stable up to the Planck scale $M_{pl}$ and also that both of the Higgs self coupling $\lambda$ and its beta function $\beta_{\lambda}$ become very small around $M_{pl}$. This fact attracts much attention, and there are many works which try to find its physical meaning \cite{Froggatt:1995rt,Froggatt:2001pa,Nielsen:2012pu,Buttazzo:2013uya,Shaposhnikov:2009pv,Meissner:2007xv,Khoze:2014xha,Kawai:2011qb,Kawai:2013wwa,Hamada:2014ofa,Hamada:2014xra,Bezrukov:2007ep,Hamada:2013mya,Hamada:2014iga,Hamada:2014wna,Hamada:2014raa,Kawamura:2013kua,Meissner:2006zh,Haba:2014sia,Hamada:2015ria,Hamada:2015wea}.

Well before the discovery of the Higgs, it was argued that the Higgs mass can be predicted to be around $130$GeV by the requirement that the minimum of the  Higgs potential becomes zero at $M_{pl}$ \cite{Froggatt:1995rt,Froggatt:2001pa}. Such a requirement (not always at $M_{pl}$) is generally called the multiple point principle (MPP). One of the good points of the MPP is its predictability: The low energy effective couplings are fixed so that the minimum of the potential vanishes. 
See \cite{Hamada:2014xka,Kawana:2014zxa,Hamada:2015fma} for example.
 
By taking the fact that the MPP can be realized in the SM into consideration, a natural question is whether such a criticality can be also realized in the models beyond the SM. One of the interesting extensions is the gauged B-L model with a classical conformality \cite{Iso:2009ss,Iso:2012jn,Okada:2010wd,Okada:2011en}. Here, ``classical conformality" means there is no mass term at the classical level without gravity. This model can be obtained by gauging the global U(1)$_{\text{B-L}}$ symmetry of the SM with the three right-handed neutrinos and a SM singlet complex scalar $\Psi$. As discussed in the following, if we neglect the Yukawa couplings between the Higgs and neutrinos, 
 there are six unknown parameters in this model. In particular, two of them are new scalar couplings: $\kappa$ and $\lambda_{\Psi}$. Therefore, in principle, these six parameters can be uniquely fixed by the MPP conditions:
\be \lambda(\Lambda_{\text{MPP}})=\lambda_{\Psi}(\Lambda_{\text{MPP}})=\kappa(\Lambda_{\text{MPP}})=\beta_{\lambda}(\Lambda_{\text{MPP}})=\beta_{\lambda_{\Psi}}(\Lambda_{\text{MPP}})=\beta_{\kappa}(\Lambda_{\text{MPP}})=0,\label{eq:con1}\e
where $\Lambda_{\text{MPP}}$ is the scale at which we impose the MPP. The analyses in this paper are based on the following assumptions:
\begin{enumerate}
\item We consider the MPP at $\Lambda_{\text{MPP}}=10^{17}$ GeV.\\
\item As well as the analyses in \cite{Iso:2009ss,Iso:2012jn}, we do not include mass terms in the Lagrangian. As a result, all the low energy scales are radiatively generated.\\
\item The Higgs mass is fixed at
\be M_{h}=125.7\text{GeV},\e
and we regard the top mass $M_{t}$ as one of the free parameters.\\
\item We assume that the small neutrino masses are produced by the seesaw mechanism via the radiative breaking of the B-L symmetry. As a result, we can neglect the Yukawa couplings $y_{\nu}$ between the Higgs and neutrinos because the typical breaking scale is very small ($\ll10^{13}$GeV).
\end{enumerate}
In subsection \ref{sub:MPP}, we will see that Eq.(\ref{eq:con1}) can be actually realized at $\Lambda_{\text{MPP}}=10^{17}$GeV.\\

One of the good features of this model is that the electroweak symmetry breaking can be triggered by the U(1)$_{\text{B-L}}$ symmetry breaking via the Coleman-Weinberg (CW) mechanism. In \cite{Iso:2012jn}, it was argued that we can naturally obtain $ v_{h}={\cal{O}}(100)$GeV by imposing $\lambda(M_{pl})=0$ and $\kappa(M_{pl})=0$. Here, the important point is that $\lambda_{\Psi}(\Lambda_{\text{MPP}})\neq0$ is needed to realize such B-L breaking\footnote{Realizing the B-L symmetry breaking when $\lambda_{\Psi}(\Lambda_{\text{MPP}})=0$ is difficult. See Section\ref{sec:b-l}.}. Therefore, if we try to combine this fact and the MPP, a natural question arises:\\
\begin{itemize}\item {\it{Is the ${\cal{O}}(100)$GeV electroweak symmetry breaking possible even if $\lambda_{\Psi}(\Lambda_{\text{MPP}})$ is small ?}}\\
\end{itemize}
In subsection 2.3, we will see that this is actually possible even if $\lambda_{\Psi}(\Lambda_{\text{MPP}})\leq 10^{-10}$. The reason for this is very simple: By tuning the parameters of the model, we can obtain the favorable scale at which U(1)$_{\text{B-L}}$ breaks so that $v_{h}$ becomes ${\cal{O}}(100)$GeV. Therefore, the B-L model is a phenomenologically very interesting model in that it can explain the natural-scale electroweak symmetry breaking while satisfying the (slightly broken) MPP. Furthermore, within this situation, we find that the mass of the B-L gauge boson is predicted to be 
\be M_{B-L}=2g_{B-L}(v_{B-L})v_{B-L}=2\sqrt{2}\times\sqrt{\frac{\lambda(v_{h})}{0.10}}\times v_{h}\simeq696\h{1mm}\text{GeV},\e 
where $v_{B-L}$ is the expectation value of $\Psi$ and we have used the typical value $\lambda(v_{h})\simeq0.1$. This is a remarkable prediction of the (slightly broken) MPP, and it is surprising that the predicted value of $M_{B-L}$ depends only on the SM parameters\footnote{Unfortunately, this value of $M_{B-L}$ is already excluded by the experiment of ATLAS \cite{Aad:2014cka}. See subsection \ref{sub:breaking}.}. 

On the other hand, there are many observational results from the cosmological side. One of the reliable possibilities to explain them is the cosmic inflation. As is well known, Higgs inflation is possible in the SM where the criticality of the Higgs potential plays an important role to realize the inflation naturally \cite{Hamada:2014wna}. Of course, such a Higgs inflation is possible in the B-L model, but, we can also consider the inflation scenario where $\Psi$ plays a roll of the inflaton \cite{Okada:2011en}. In this paper, we study the $\lambda_{\Psi}\Psi^{4}$ inflation with the non-minimal gravitational coupling $\xi\Psi^{2}{\cal{R}}$. Our analysis is based on the following condition:
\begin{itemize}\item {\it{We consider the inflation under the situation where the minimum of the Higgs potential vanishes at $\Lambda_{\text{MPP}}=10^{17}$GeV and the  electroweak symmetry breaking occurs at ${\cal{O}}(100)$GeV.}}
\end{itemize}
In the following discussion, we will see that this condition strongly constrains the parameters, and as a result, we can obtain the unique cosmological predictions\footnote{Here, we use ``unique" in the sense that our predictions do not so depend on the parameters of the model except for $\lambda_{\Psi}$, $\xi$ and the initial value of $\Psi$.} which are consistent with the recent observed values by Planck \cite{Planck:2015xua} and BICEP2 \cite{Ade:2014xna}. 
\\
\\
This paper is organized as follows. In Section \ref{sec:b-l}, we study the MPP and the B-L symmetry breaking from the point of view of the slightly broken MPP. In Section \ref{sec:inf}, we investigate the inflation scenario where the SM singlet complex scalar $\Psi$ plays a roll of the inflaton. In Section \ref{sec:sum}, we give summary.

\section{MPP of the B-L Model and Symmetry Breaking}\label{sec:b-l}
\begin{table}
\begin{center}
\begin{tabular}{|c|c|c|c|c|}\hline
                   & $SU(3)_{c}$ & $SU(2)_{L}$ & $U(1)_{Y}$ & $U(1)_{B-L}$ \\ \hline
$Q^{i}_{L}$ & \bf{3} & \bf{2} & $+1/6$ & $+1/3$\\ \hline
$u^{i}_{R}$ & \bf{3} & \bf{1} & $+2/3$ & $+1/3$\\ \hline
$d^{i}_{R}$ & \bf{3} & \bf{1} & $-1/3$ & $+1/3$\\ \hline
$\ell^{i}_{L}$ & \bf{1} & \bf{2} & $-1/2$ & $-1$\\ \hline
$\nu^{i}_{R}$ & \bf{1} & \bf{1} & $0$ & $-1$\\ \hline
$e^{i}_{R}$ & \bf{1} & \bf{1} & $-1$ & $-1$\\ \hline
$H$            & \bf{1} & \bf{2} & $-1/2$ & $0$\\ \hline
$\Psi$         & \bf{1} & \bf{1} & $0$ & $+2$\\ \hline

\end{tabular}
\end{center}
\caption{The particle contents of the B-L model and their charges except for the gauge bosons. Here, $i$ represents the generation.}
\end{table} 
The flow of this section is as follows. In subsection \ref{sub:rev}, we shortly review the gauged B-L model. In subsection \ref{sub:MPP}, we consider the MPP of this model. 
In subsection \ref{sub:breaking}, we study whether the ${\cal{O}}(100)$GeV electroweak symmetry breaking can be realized even if $\lambda_{\Psi}\left(\Lambda_{\text{MPP}}\right)$ is very small.

\subsection{Short Review of the B-L Model}\label{sub:rev}
In this subsection, we briefly review the B-L extension of the SM. Here, our discussion is mainly based on \cite{Chankowski:2006jk}. As mentioned in the introduction, this model can be obtained by gauging the global U(1)$_{\text{B-L}}$ symmetry. The kinetic terms of the two U(1) gauge fields are given as follows:
\be {\cal{L}}_{kin}=-\frac{1}{4}F^{\mu\nu}F_{\mu\nu}-\frac{1}{4}F_{\text{B-L}}^{\mu\nu}F_{\text{B-L}\mu\nu}-\frac{\omega}{4}F_{\text{B-L}}^{\mu\nu}F_{\mu\nu},\e
where $\omega (\in \mathbb{R})$ represents the kinetic mixing. The U(1) part of the covariant derivative of a matter field $\phi_{k}$ is given by
\be {\cal{D}}_{\mu}=\partial_{\mu}+i\sum_{i=1}^{2}\sum_{j=1}^{2}Y_{k}^{i}g_{ij}A_{\mu}^{j},\label{eq:cov}\e
where $A_{\mu}^{1}$ and $A_{\mu}^{2}$ are the gauge fields of U(1)$_{Y}$ and U(1)$_{\text{B-L}}$ respectively, $Y_{k}^{i}$ are the U(1) charges and $g_{ij}$ represent the U(1) gauge couplings. We can remove the mixing term by changing $A_{\mu}^{1}$ and $A_{\mu}^{2}$ to the new fields $A_{\mu}^{Y}$ and $A_{\mu}^{B-L}$:
\be A_{\mu}^{1}=\frac{1}{\sqrt{2(1+\omega)}}A_{\mu}^{Y}+\frac{1}{\sqrt{2(1-2\omega)}}A_{\mu}^{B-L}\h{2mm},\h{2mm} A_{\mu}^{2}=\frac{1}{\sqrt{2(1+\omega)}}A_{\mu}^{Y}-\frac{1}{\sqrt{2(1-2\omega)}}A_{\mu}^{B-L}.\label{eq:rot1}\e
We simply express Eq.(\ref{eq:rot1}) as $A_{\mu}^{i}=\sum_{\alpha}R^{i}_{\alpha}A_{\mu}^{\alpha}$. By this transformation,  the new gauge couplings are 
\be g'_{i\alpha}:=\sum_{j}g_{ij}R^{j}_{\alpha}.\e
We denote $g'_{i\alpha}$ as $g_{YY}$, $g_{YE}$, $g_{EY}$ and $g_{EE}$ without a prime in the following discussion. Only three of them are meaningful because we can further rotate the gauge fields without producing the mixing term: 
\[\left(\begin{array}{lcr} A^{Y}\\ A^{B-L}\end{array}\right)=\left(\begin{array}{lcr} \cos\theta && -\sin\theta \\ \sin\theta && \cos\theta\end{array}\right)\left(\begin{array}{lcr} \tilde{A}^{Y}\\ \tilde{A}^{B-L}\end{array}\right).\]
Thus, we can choose the angle $\theta$ so that one of $g_{\alpha\beta}$ vanishes. For convenience, we take the following bases: 
\be B_{\mu}:=\frac{g_{EE}A_{\mu}^{B-L}+g_{EY}A^{Y}_{\mu}}{\sqrt{g_{EE}^{2}+g_{EY}^{2}}}\h{2mm},\h{2mm}E_{\mu}:=\frac{-g_{EY}A_{\mu}^{B-L}+g_{EE}A^{Y}_{\mu}}{\sqrt{g_{EE}^{2}+g_{EY}^{2}}}.\e
In this bases, the second term of Eq.(\ref{eq:cov}) becomes
\be g_{Y}Y^{Y}_{k}B_{\mu}+(g_{B-L}Y^{B-L}_{k}+g_{\text{mix}}Y^{Y}_{k})E_{\mu},\label{eq:form}\e
where
\be g_{Y}:=\frac{g_{EE}g_{YY}-g_{EY}g_{YE}}{\sqrt{g_{EE}^{2}+g_{EY}^{2}}}\h{2mm},\h{2mm}g_{B-L}:=\sqrt{g_{EE}^{2}+g_{EY}^{2}}\h{2mm},\h{2mm}g_{\text{mix}}:=\frac{g_{YE}g_{EE}+g_{EY}g_{YY}}{\sqrt{g_{EE}^{2}+g_{EY}^{2}}}.\e
As a result, $B_{\mu}$ plays a roll of the ordinary U(1)$_{Y}$ gauge field, and $E_{\mu}$ is a new gauge field which can have a mass if the B-L symmetry is broken. We use Eq.(\ref{eq:form}) for the calculations of the RGEs in \ref{app:beta}.

The particle contents (except for the gauge bosons) and their charges are presented in Table1. In addition to the SM particles, there are three right-handed neutrinos and a SM singlet complex scalar whose U(1)$_{\text{B-L}}$ charge is $+2$.  The relevant  terms of the renormalizable Lagrangian are
\begin{align} {\cal{L}}\supset -\lambda\left(H^{\dagger}H\right)^{2}-\lambda_{\Psi}\left(\Psi^{\dagger}\Psi\right)^{2}-\kappa\left(H^{\dagger}H\right)\left(\Psi^{\dagger}\Psi\right)\nonumber\\
-\sum_{ij}y^{ij}_{\nu}\bar{\nu}_{R}^{i}H^{\dagger} \ell_{L}^{j}-\frac{1}{2}\sum_{ij}Y^{ij}_{R}\bar{\nu^{c}}_{R}^{i}\nu_{R}^{j}\Psi+\text{h.c}.
\end{align}
In the following discussion, we use the bases such that $y_{\nu}^{ij}$ and $Y_{R}^{ij}$ are real and diagonalized, and assume that they are equal respectively for the three generations. 
As a result, by including the top mass $M_{t}$, there are seven unknown parameters in this model:
\be M_{t}\h{2mm},\h{2mm}g_{B-L}\h{2mm},\h{2mm}g_{\text{mix}}\h{2mm},\h{2mm}\lambda_{\Psi}\h{2mm},\h{2mm}\kappa\h{2mm},\h{2mm}y_{\nu}\h{2mm},\h{2mm}Y_{R}.\e
If we assume that the small neutrino masses ($\lesssim$ 1eV) are generated by the ordinary seesaw mechanism triggered by the U(1)$_{\text{B-L}}$ symmetry breaking at a low energy scale ($\ll10^{13}$ GeV), $y_{\nu}$ should be very small, and its effects to the RGEs can be negligible. In this paper, we assume such a  situation.

\subsection{Multiple Point Principle}\label{sub:MPP}
To understand how these couplings behave at a high energy scale, we need to know the RGEs. The two-loop RGEs of this model are presented in \ref{app:beta}. Furthermore, the one-loop effective potentials in the Landau gauge are as follows\footnote{Here, we neglect the one-loop contributions which include $\lambda$, $\lambda_{\Psi}$ and $\kappa$ because their effects are very small when we consider the MPP.}:
\be V^{H}_{\text{eff}}(\mu,\phi)=\frac{\lambda(\mu)}{4}\phi^{4}+V^{H}_{\text{1loop}}(\mu,\phi)\h{2mm},\h{2mm}V^{\Psi}_{\text{eff}}(\mu,\Psi)=\frac{\lambda_{\Psi}(\mu)}{4}\Psi^{4}+V^{\Psi}_{\text{1loop}}(\mu,\Psi),\e
\begin{align} V^{H}_{\text{1loop}}&(\mu,\phi):=e^{4\Gamma(\mu)}\Biggl\{-12\cdot\frac{M_{t}(\phi)^{4}}{64\pi^{2}}\left[\log\left(\frac{M_{t}(\phi)^{2}}{\mu^{2}}\right)-\frac{3}{2}+2\Gamma(\mu)\right] \nonumber\\
&+6\cdot\frac{M_{W}(\phi)^{4}}{64\pi^{2}}\left[\log\left(\frac{M_{W}(\phi)^{2}}{\mu^{2}}\right)-\frac{5}{6}+2\Gamma(\mu)\right]+3\cdot\frac{M_{Z}(\phi)^{4}}{64\pi^{2}}\left[\log\left(\frac{M_{Z}(\phi)^{2}}{\mu^{2}}\right)-\frac{5}{6}+2\Gamma(\mu)\right]\Biggl\},\nonumber\end{align}
\begin{align} V^{\Psi}_{\text{1loop}}(\mu,\Psi):=e^{4\Gamma_{\Psi}(\mu)}\Biggl\{-6\cdot\frac{M_{R}(\Psi)^{4}}{64\pi^{2}}&\left[\log\left(\frac{M_{R}(\Psi)^{2}}{\mu^{2}}\right)-\frac{3}{2}+2\Gamma_{\Psi}(\mu)\right] \nonumber\\
&+3\cdot\frac{M_{B-L}(\Psi)^{4}}{64\pi^{2}}\left[\log\left(\frac{M_{B-L}(\Psi)^{2}}{\mu^{2}}\right)-\frac{5}{6}+2\Gamma_{\Psi}(\mu)\right]\Biggl\}
,\end{align}
where
\begin{align} M_{t}(\phi)=\frac{y_{t}(\mu)}{\sqrt{2}}&\phi\h{4mm},\h{3mm}M_{W}(\phi)=\frac{g_{2}(\mu)}{2}\phi\h{2mm},\h{2mm}M_{Z}(\phi)=\frac{\sqrt{g_{2}^{2}(\mu)+g_{\text{Y}}^{2}(\mu)}}{2}\phi,\nonumber\\
&M_{R}(\Psi)=\frac{Y_{R}(\mu)}{\sqrt{2}}\Psi
\h{2mm},\h{2mm}M_{B-L}(\Psi)^{2}=2^{2}g_{B-L}(\mu)^{2}\Psi^{2}.\end{align}
Here, $\mu$ is the renormalization scale and $\Gamma$, $\Gamma_{\Psi}$ are the wave function renormalizations. To minimize the one-loop contributions, we take $\mu=\phi\h{1mm}(\Psi)$ in the following discussion\footnote{Precisely speaking, $\mu$ should be determined as a function of $\phi$ and $\Psi$ by minimizing the one-loop effective potential. However, in this paper, we simply choose $\mu=\phi \h{1mm}(\Psi)$ when we focus on $\lambda^{\text{eff}}\h{1mm}(\lambda_{\Psi}^{\text{eff}})$. It is known that this choice is a good approximation \cite{Hamada:2014wna}.}. From these results, we can define the effective self couplings and their effective beta functions as follows:
\be \lambda^{\text{eff}}(\phi):=\frac{4V^{H}_{\text{eff}}(\phi)}{\phi^{4}}\h{2mm},\h{2mm}\beta^{\text{eff}}_{\lambda}:=\frac{d\lambda^{\text{eff}}(\phi)}{d\ln\phi},\e
\be\lambda^{\text{eff}}_{\Psi}(\Psi):=\frac{4V^{\Psi}_{\text{eff}}(\Psi)}{\Psi^{4}}\h{2mm},\h{2mm}\beta^{\text{eff}}_{\lambda_{\Psi}}:=\frac{d\lambda_{\Psi}^{\text{eff}}(\Psi)}{d\ln\Psi}.\e
Fig.\ref{fig:typical} shows the typical behaviors of $\lambda^{\text{eff}}(\phi)$ and its parameter dependences. Here, for the later convenience, the initial values of $\lambda_{\Psi}$, $\kappa$, $g_{B-L}$, $g_{\text{mix}}$ and $Y_{R}$ are given at $\Lambda_{\text{MPP}}=10^{17}$GeV, and their typical values are chosen to be $0.1$ respectively. One can see that $\lambda^{\text{eff}}(\phi)$ depends weakly on $g_{B-L}$ and $Y_{R}$ because they appear in $\beta_{\lambda}$ at two-loop level.

\begin{figure}
\begin{center}
\begin{tabular}{c}
\begin{minipage}{0.5\hsize}
\begin{center}
\includegraphics[width=9cm]{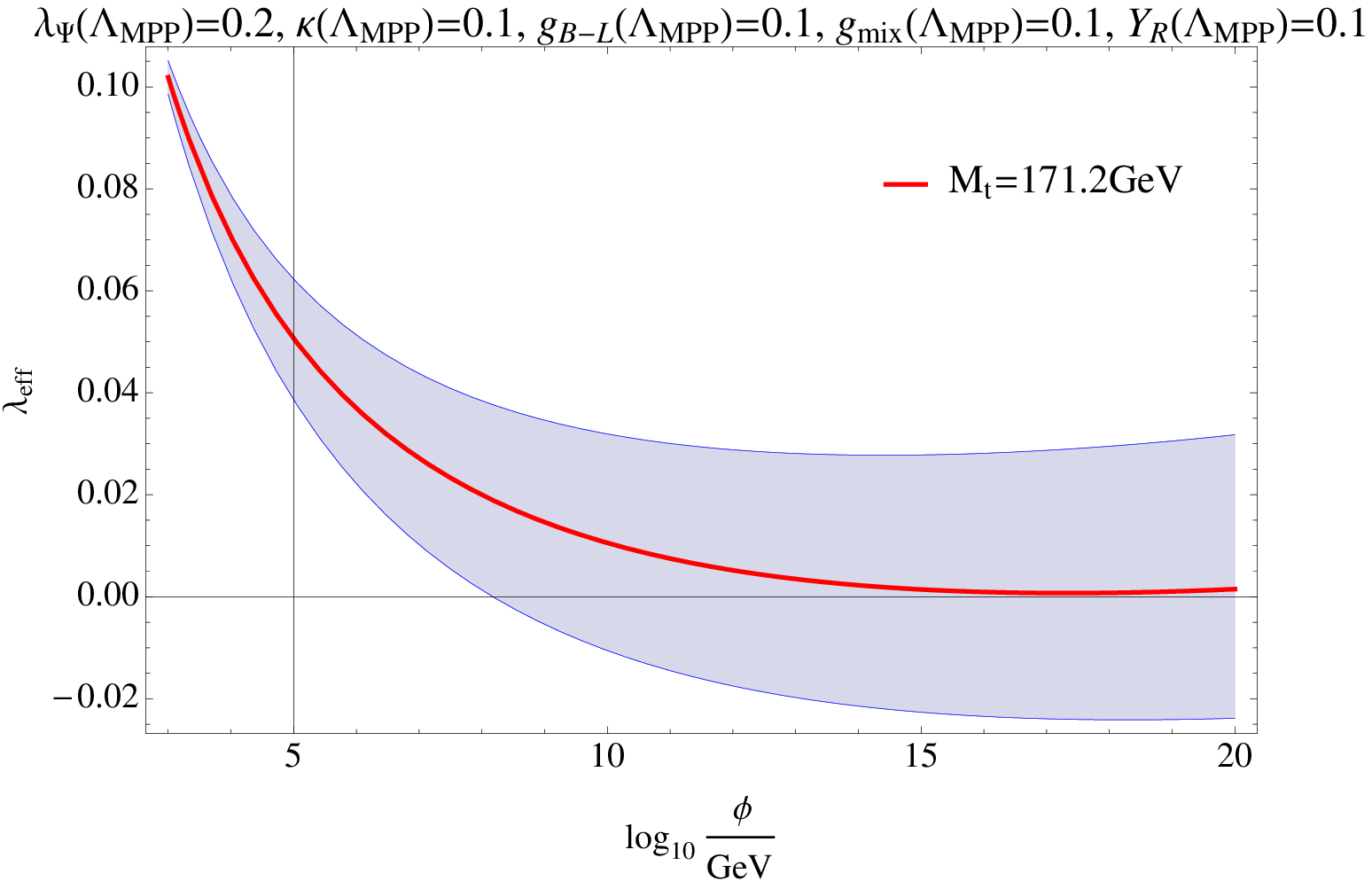}
\end{center}
\end{minipage}
\begin{minipage}{0.5\hsize}
\begin{center}
\includegraphics[width=9cm]{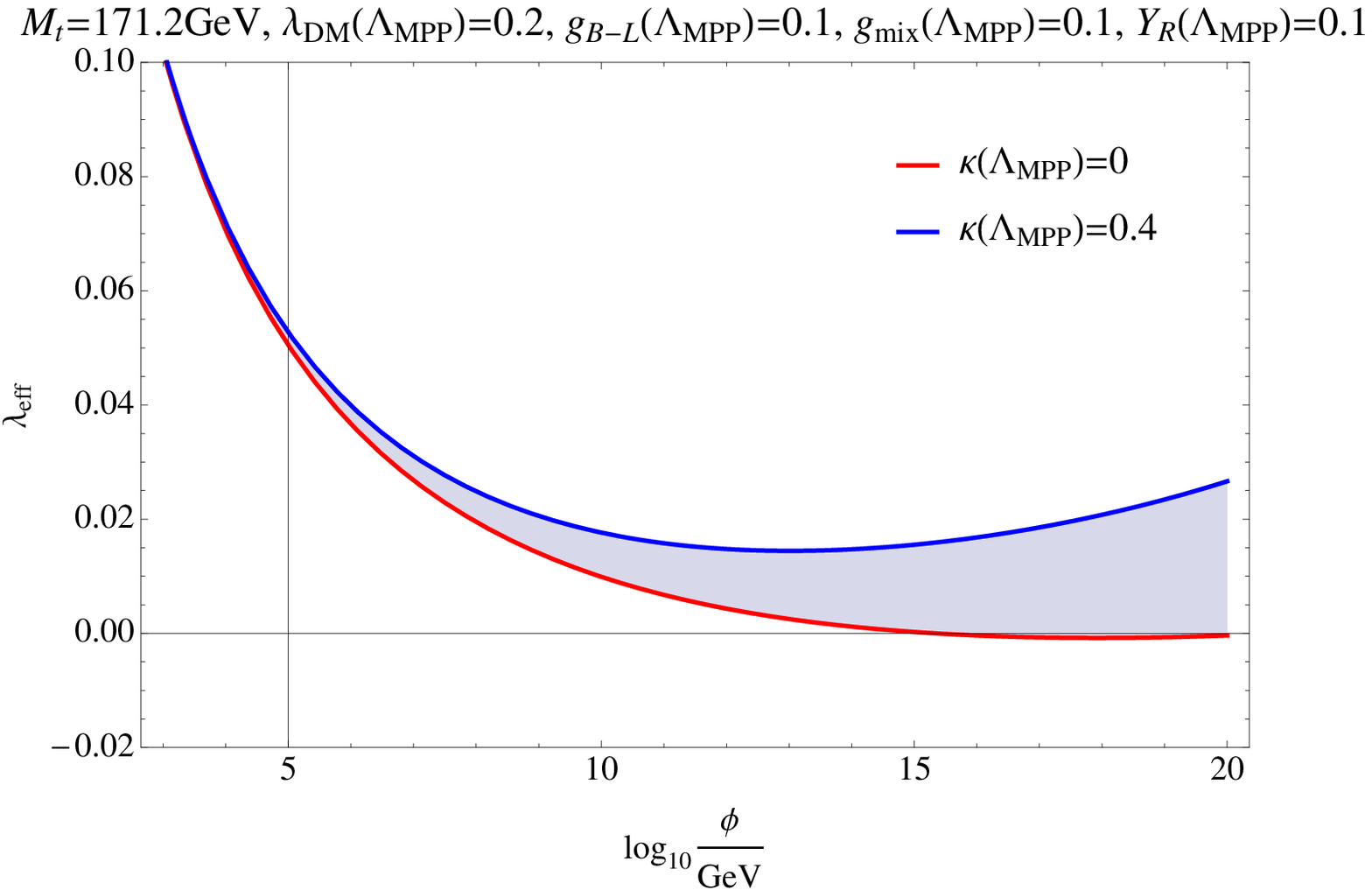}
\end{center}
\end{minipage}
\\
\\
\begin{minipage}{0.5\hsize}
\begin{center}
\includegraphics[width=9cm]{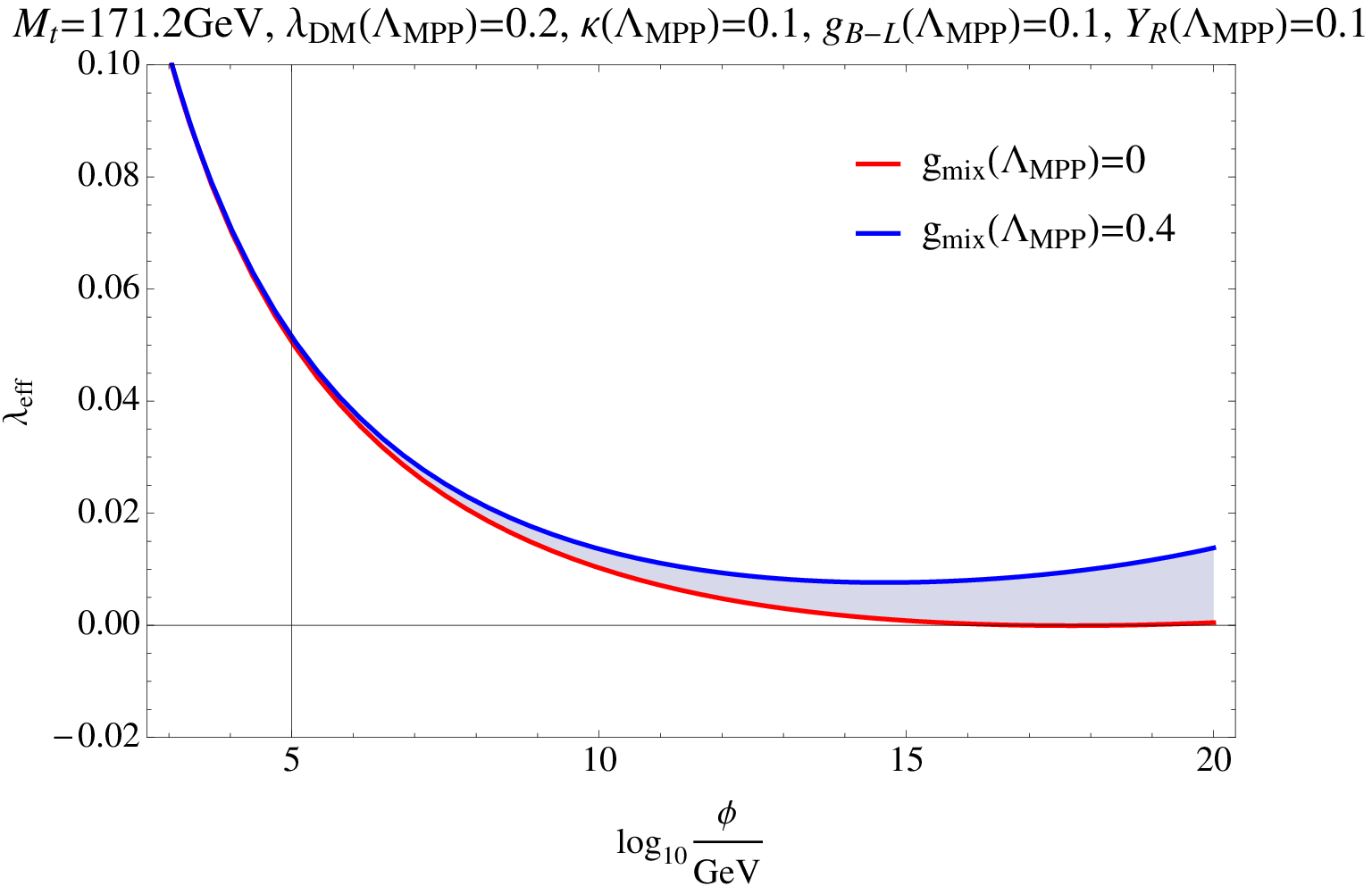}
\end{center}
\end{minipage}
\begin{minipage}{0.5\hsize}
\begin{center}
\includegraphics[width=9cm]{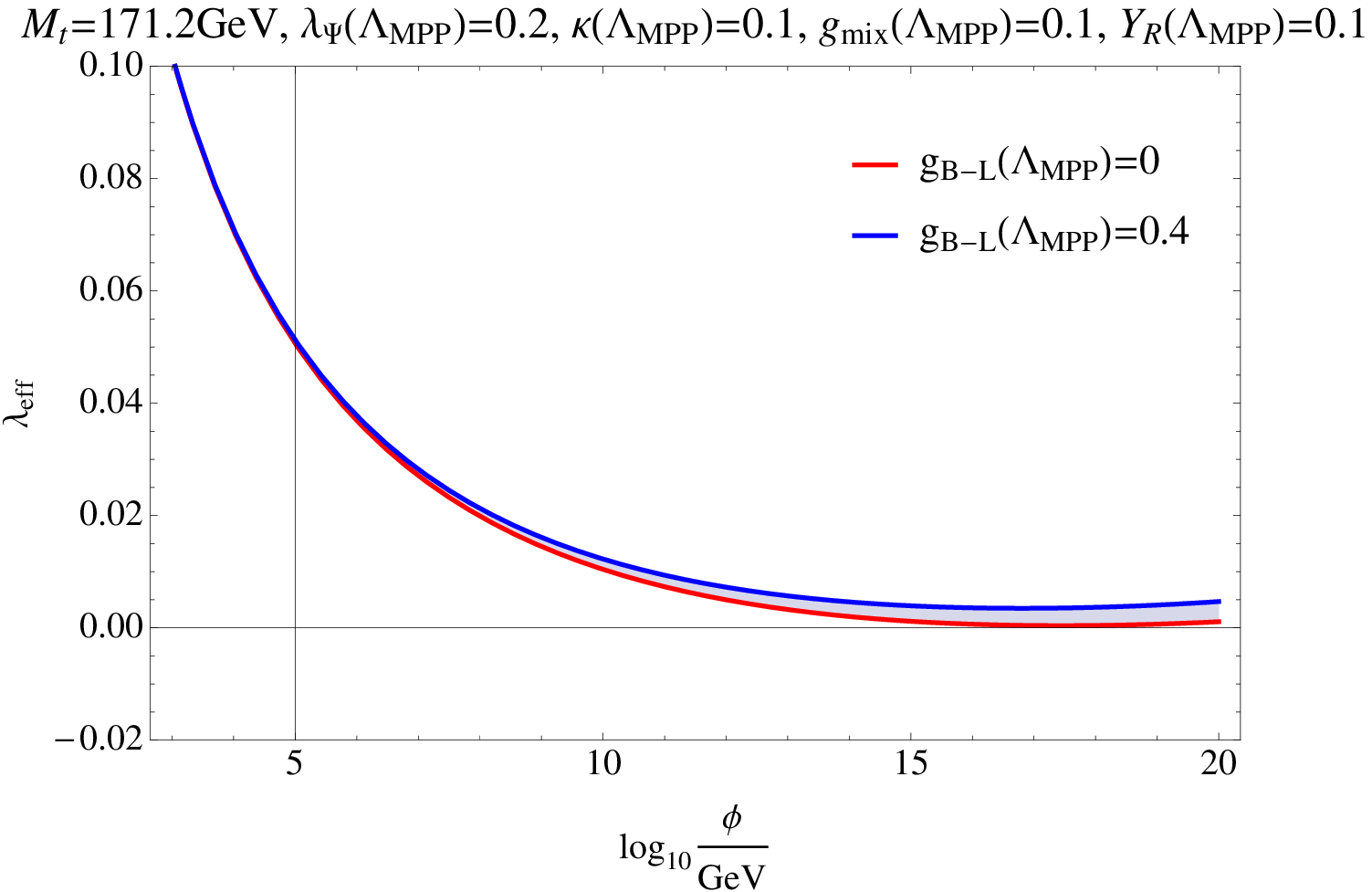}
\end{center}
\end{minipage}
\\
\\
\begin{minipage}{0.5\hsize}
\begin{center}
\includegraphics[width=9cm]{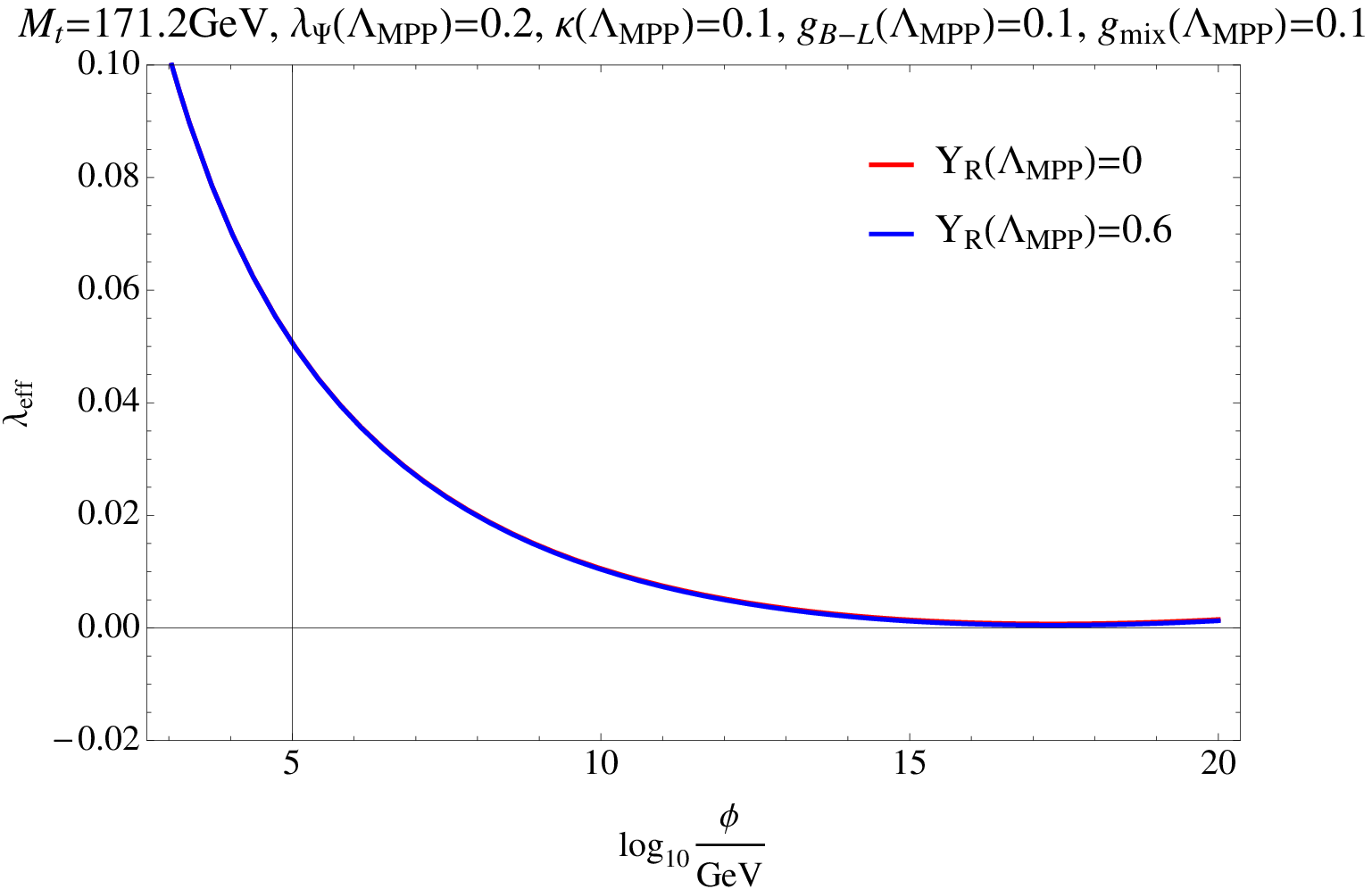}
\end{center}
\end{minipage}
\begin{minipage}{0.5\hsize}
\begin{center}
\includegraphics[width=9cm]{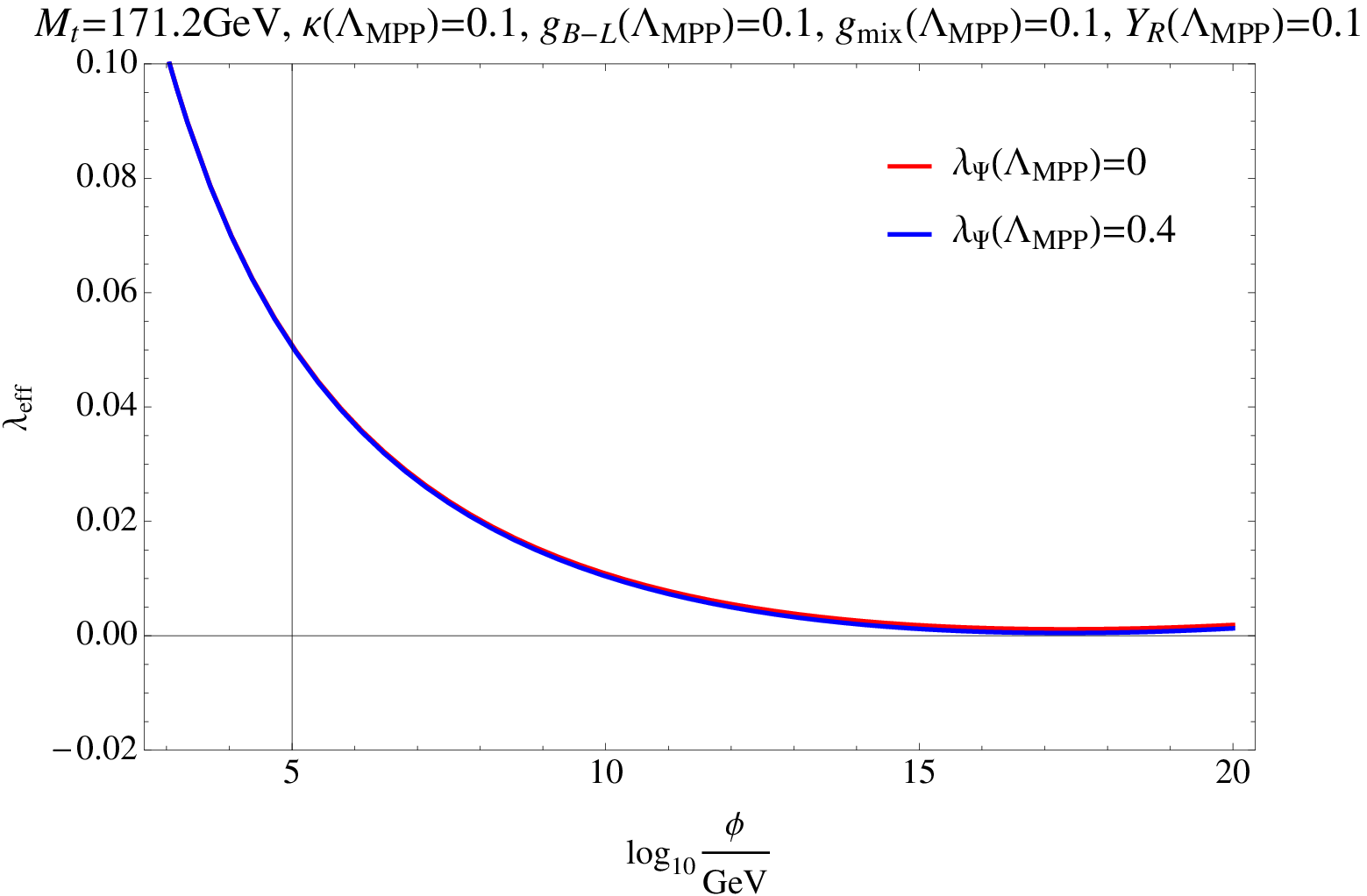}
\end{center}
\end{minipage}
\end{tabular}
\end{center}
\caption{The runnings of the Higgs effective self coupling $\lambda_{\text{eff}}$ as a function of $\phi$. The upper left (right) panel shows the $M_{t}\h{1mm}(\kappa(\Lambda_{\text{MPP}}))$ dependence. In the case of $M_{t}$, the blue band corresponds 95$\%$ CL deviation from $171.2$GeV. The middle left (right) panel shows the $g_{\text{mix}}(\Lambda_{\text{MPP}})\h{1mm}(g_{B-L}(\Lambda_{\text{MPP}})))$ dependence. The lower left (right) panel shows the $Y_{R}(\Lambda_{\text{MPP}})\h{1mm}(\lambda_{\Psi}(\Lambda_{\text{MPP}}))$ dependence.}
\label{fig:typical}
\end{figure}
\newpage

\begin{figure}
\begin{center}
\begin{tabular}{c}
\begin{minipage}{0.5\hsize}
\begin{center}
\includegraphics[width=8.5cm]{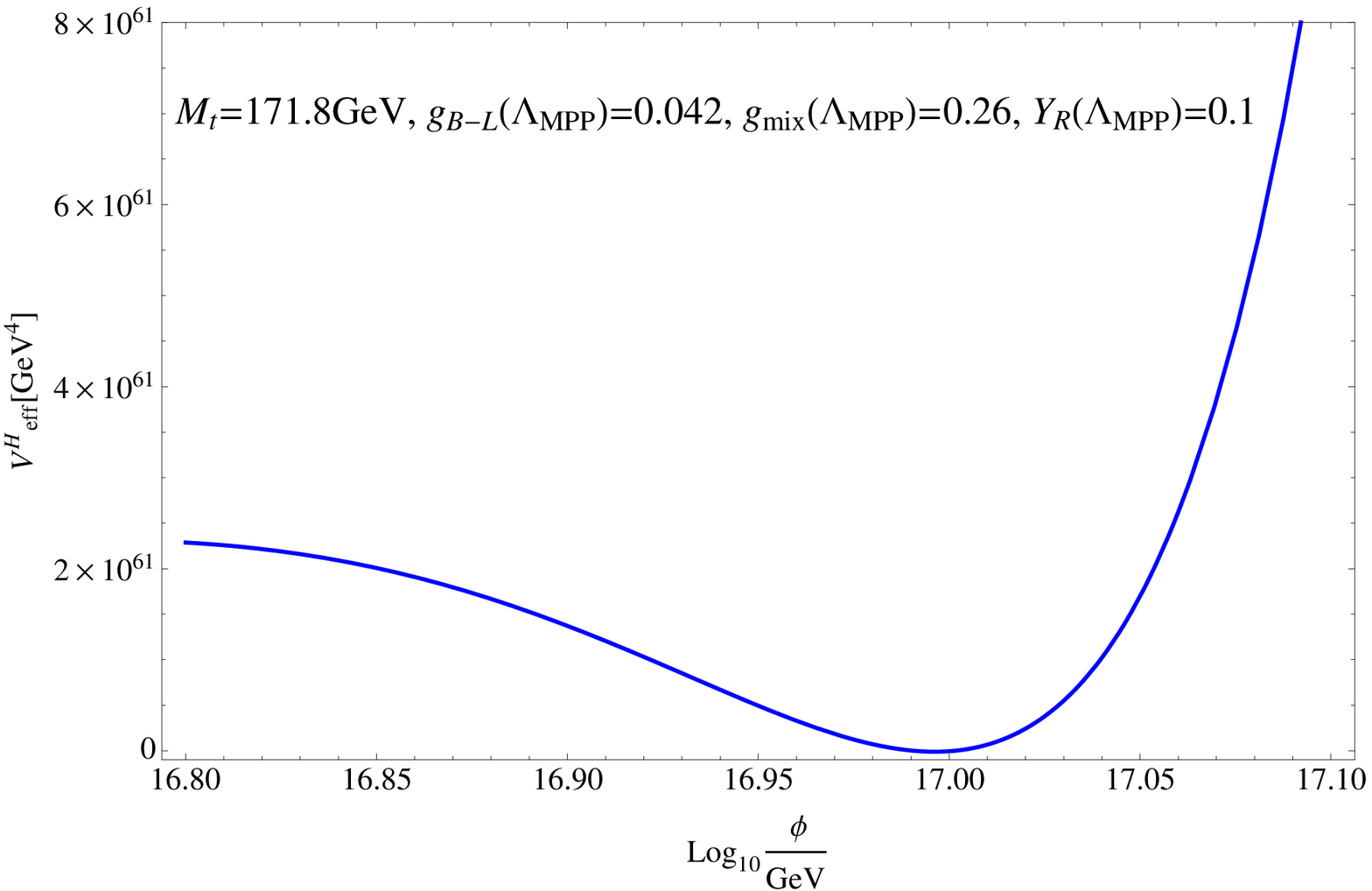}
\end{center}
\end{minipage}
\begin{minipage}{0.5\hsize}
\begin{center}
\includegraphics[width=8.5cm]{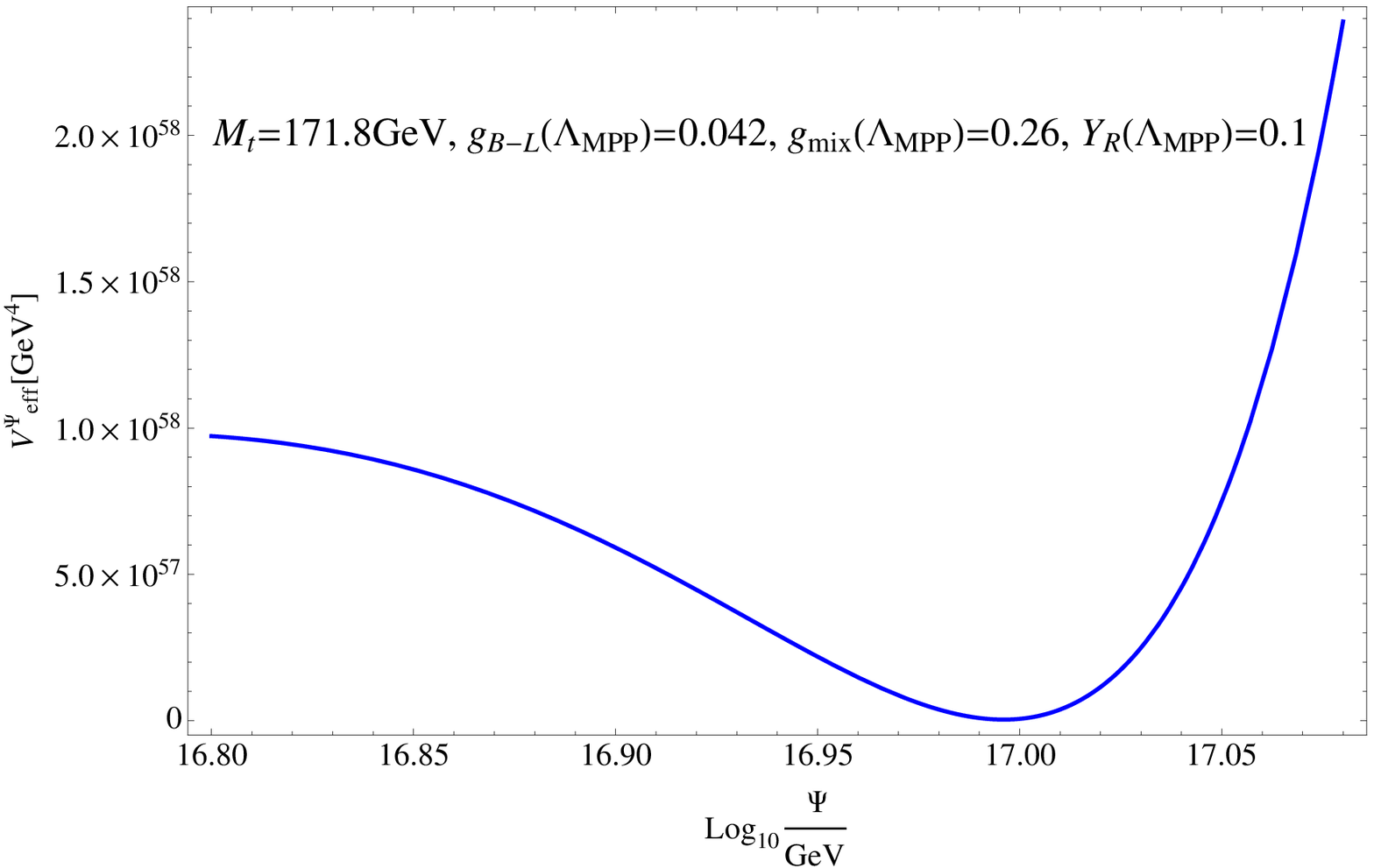}
\end{center}
\end{minipage}
\\
\\
\begin{minipage}{0.5\hsize}
\begin{center}
\includegraphics[width=8.5cm]{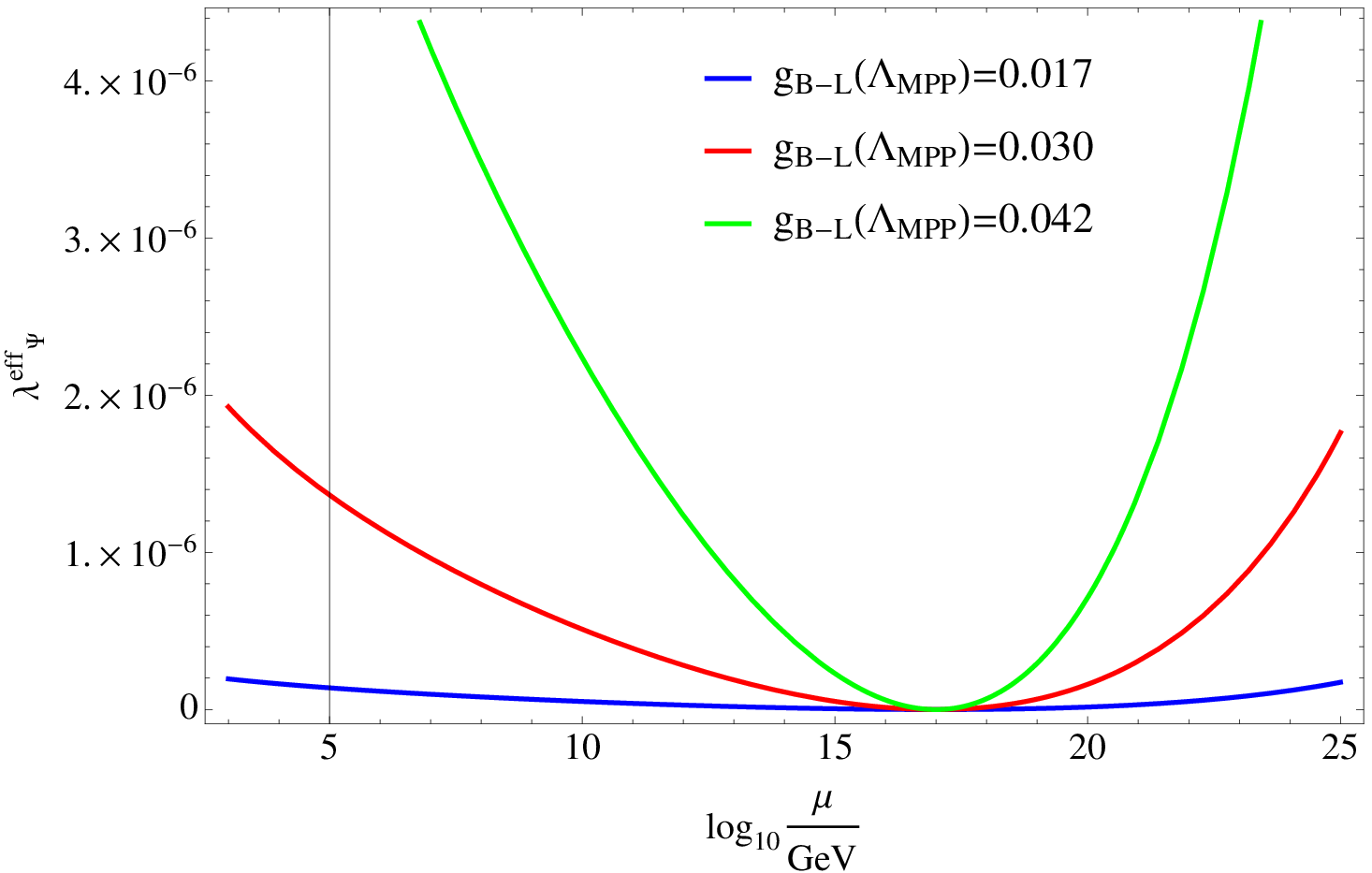}
\end{center}
\end{minipage}
\begin{minipage}{0.5\hsize}
\begin{center}
\includegraphics[width=8.5cm]{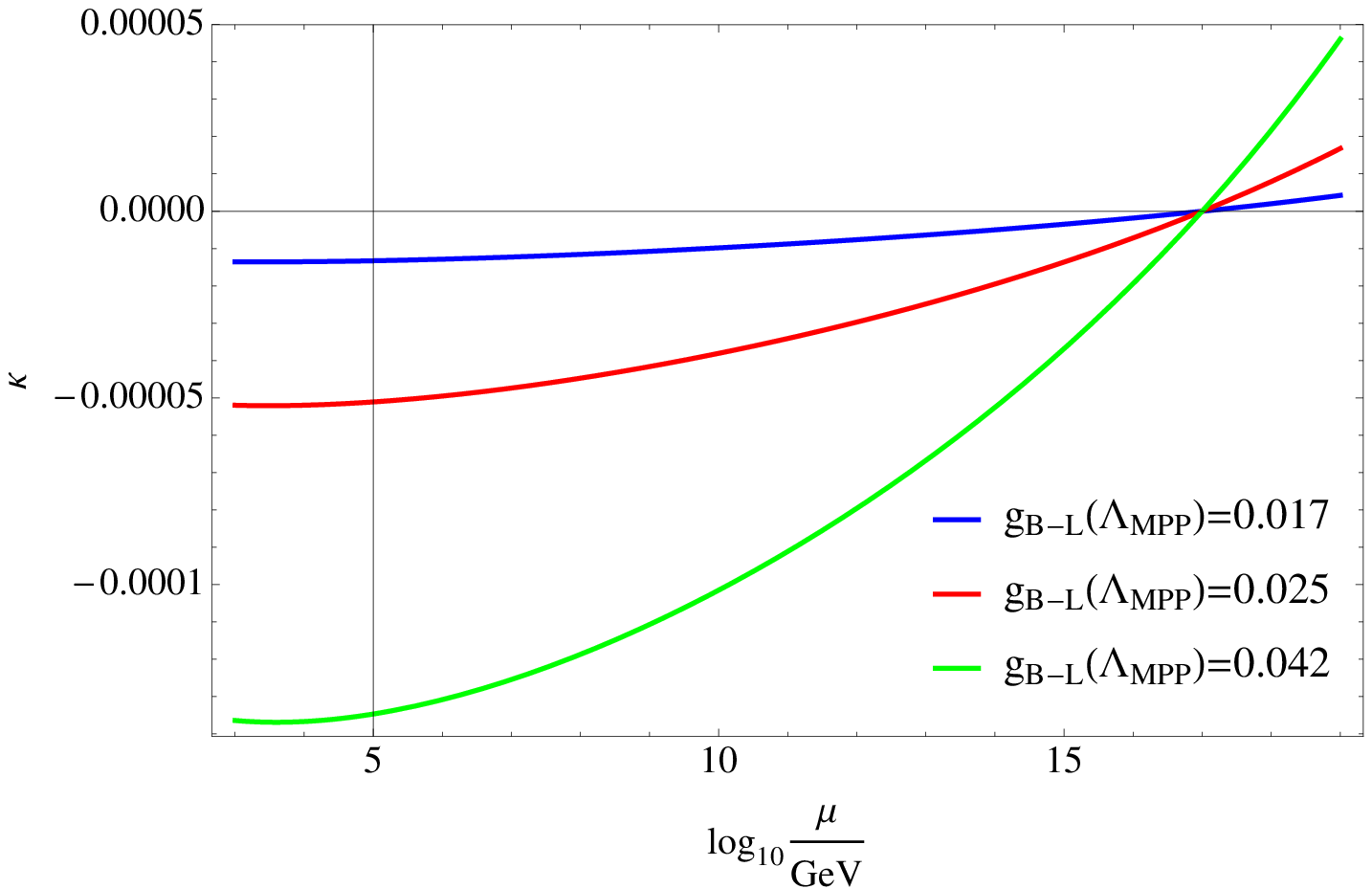}
\end{center}
\end{minipage}
\end{tabular}
\end{center}
\caption{The effective potentials (upper) and the runnings of $\lambda_{\Psi}^{\text{eff}}$ and $\kappa$ which satisfy the MPP conditions (lower). The upper left (right) panel shows $V_{\text{eff}}^{H}\h{1mm}(V_{\text{eff}}^{\Psi})$. They are exactly flat at $\Lambda_{\text{MPP}}=10^{17}$GeV. In the lower panels, we leave $g_{B-L}(\Lambda_{\text{MPP}})$ as a free parameter. The different colors correspond to the different values of $g_{B-L}(\Lambda_{\text{MPP}})$.  }
\label{fig:critical pot}
\end{figure}

\noindent Now, let us consider the MPP. By including the top mass $M_{t}$ and neglecting $y_{\nu}$, there are six parameters in this model:
\be M_{t}\h{2mm},\h{2mm}g_{B-L}\h{2mm},\h{2mm}g_{\text{mix}}\h{2mm},\h{2mm}\lambda_{\Psi}\h{2mm},\h{2mm}\kappa\h{2mm},\h{2mm}Y_{R}.\label{eq:para}\e
Therefore, in principle, they are uniquely determined by the MPP conditions:
\be \lambda^{\text{eff}}(\Lambda_{\text{MPP}})=\lambda^{\text{eff}}_{\Psi}(\Lambda_{\text{MPP}})=\kappa(\Lambda_{\text{MPP}})=\beta_{\lambda}^{\text{eff}}(\Lambda_{\text{MPP}})=\beta_{\lambda_{\Psi}}^{\text{eff}}(\Lambda_{\text{MPP}})=\beta_{\kappa}(\Lambda_{\text{MPP}})=0.\label{eq:MPP}\e
Among them, $\lambda^{\text{eff}}_{\Psi}(\Lambda_{\text{MPP}})=\kappa(\Lambda_{\text{MPP}})=0$ are just the initial conditions of $\lambda_{\Psi}$ and $\kappa$, and other conditions give us constraints between the remaining parameters. We can understand such constraints qualitatively from the one-loop RGEs: 

\begin{itemize}\item $\beta_{\lambda}^{\text{eff}}(\Lambda_{\text{MPP}})=0$ mainly relates $M_{t}$ and $g_{\text{mix}}$ because they appear in $\beta_{\lambda}$ at one-loop level (see Eq.(\ref{eq:betalam}) in \ref{app:beta}). As a result, we can fix $M_{t}$ and $g_{\text{mix}}$ by $\lambda^{\text{eff}}(\Lambda_{\text{MPP}})=\beta_{\lambda}^{\text{eff}}(\Lambda_{\text{MPP}})=0$. They are
\be 171.74\h{1mm}\text{GeV}\leq M_{t} \leq 171.82\h{1mm}\text{GeV}\h{2mm},\h{2mm}0.21\leq g_{\text{mix}}(\Lambda_{\text{MPP}})\leq0.27\e
according to $0\leq g_{B-L}(\Lambda_{\text{MPP}})\leq 0.4$ \footnote{$Y_{R}(\Lambda_{\text{MPP}})$ dependence is negligible.}.\\
\item We can obtain a relation between $g_{B-L}(\Lambda_{\text{MPP}})$ and $Y_{R}(\Lambda_{\text{MPP}})$ by $\beta_{\lambda_{\Psi}}(\Lambda_{\text{MPP}})=0$ because the one-loop part of $\beta_{\lambda_{\Psi}}$ at $\Lambda_{\text{MPP}}$ is
\be \beta_{\lambda_{\Psi}}|_{\text{1-loop}}(\Lambda_{\text{MPP}})=\frac{1}{16\pi^{2}}\left(96g_{B-L}^{4}-3Y_{R}^{4}\right).\label{eq:1beta}\e
\item Finally, $g_{B-L}(\Lambda_{\text{MPP}})$ (or $Y_{R}(\Lambda_{\text{MPP}})$) can be fixed at 0 by $\beta_{\kappa}(\Lambda_{\text{MPP}})=0$ because the one-loop part of $\beta_{\kappa}$ at $\Lambda_{\text{MPP}}$ is
\be \beta_{\kappa}|_{\text{1-loop}}(\Lambda_{\text{MPP}})=\frac{1}{16\pi^{2}}\left(12g_{B-L}^{2}g_{\text{\text{mix}}}^{2}-12Y_{R}^{2}y_{\nu}^{2}\right)\simeq \frac{12g_{B-L}^{2}g_{\text{mix}}^{2}}{16\pi^{2}}.\e
\end{itemize}
In Fig.\ref{fig:critical pot}, we show the effective potentials (upper) and the runnings (lower) of $\lambda_{\Psi}^{\text{eff}}$ and $\kappa$ which satisfy the above MPP conditions. Here, in the lower panels, we leave $g_{B-L}(\Lambda_{\text{MPP}})$ as a free parameter. 
One can see that the flat potentials can be actually realized at $\Lambda_{\text{MPP}}$.
\\
\\
{\it{\textbf{Summary}}} : From the MPP at $\Lambda_{\text{MPP}}=10^{17}$GeV, the parameters of the gauged B-L \\ \h{2.5cm}extension of the SM are fixed at
\begin{align} &M_{t}\simeq171.8\text{GeV}\h{2mm},\h{2mm}g_{B-L}(\Lambda_{\text{MPP}})\simeq0\h{2mm},\h{2mm}g_{\text{mix}}(\Lambda_{\text{MPP}})\simeq0.2\h{2mm},\h{2mm}\nonumber\\
&\lambda_{\Psi}(\Lambda_{\text{MPP}})\simeq0\h{2mm},\h{2mm}\kappa(\Lambda_{\text{MPP}})\simeq0\h{2mm},\h{2mm}Y_{R}(\Lambda_{\text{MPP}})\simeq0.\end{align}

\subsection{Electroweak Symmetry Breaking by Breaking the MPP}\label{sub:breaking}
We first explain how the electroweak symmetry breaking is triggered by the B-L symmetry breaking. If $\Psi$ has an expectation value $\langle\Psi\rangle:=v_{B-L}/\sqrt{2}$, the interaction term $-\kappa\left(H^{\dagger}H\right)(\Psi^{\dagger}\Psi)$ produces the mass term of $H$:
\be {\cal{L}}\ni-\frac{\kappa}{2}v_{B-L}^{2}H^{\dagger}H.\e
Thus, if $\kappa$ is negative at the B-L breaking scale, the electroweak symmetry breaking occurs, and the corresponding Higgs expectation value $v_{h}$ is given by
\be v_{h}=\sqrt{-\frac{\kappa}{2\lambda}}\times v_{B-L}\Biggl{|}_{\mu=v_{h}}.\label{eq:vev}\e
This is a relation between $v_{h}$ and $v_{B-L}$. We must consider a few questions to realize the electroweak symmetry breaking at ${\cal{O}}(100)$GeV:
\\
\\
{\it{\textbf{Question1}}} :  Does the B-L symmetry breaking actually occur? Especially, is it possible to realize it under the situation where the MPP is exactly satisfied ? 
\\
\\
See the lower left panel of Fig.\ref{fig:critical pot} once again. This shows the running of $\lambda_{\Psi}^{\text{eff}}$ when the MPP conditions are satisfied. One can see that $\lambda_{\Psi}^{\text{eff}}$ is a monotonically decreasing function in the $\mu\leq\Lambda_{\text{MPP}}$ region. Thus, we can not obtain the B-L symmetry breaking if the MPP is realized exactly. However, as discussed in \cite{Iso:2012jn}, the situation changes when $\lambda_{\Psi}^{\text{eff}}(\Lambda_{\text{MPP}})>0$ and $\beta_{\lambda_{\Psi}}^{\text{eff}}(\Lambda_{\text{MPP}})>0$, which mean the breaking of the MPP.  See the upper and middle left panels of Fig.\ref{fig:Psi}. They show the runnings of $\lambda_{\Psi}^{\text{eff}}$ when $\lambda_{\Psi}^{\text{eff}}(\Lambda_{\text{MPP}})=10^{-10}$ and $10^{-12}$ respectively\footnote{In Section \ref{sec:inf}, we will see that $\lambda_{\Psi}^{\text{eff}}$ is required to be small to explain the cosmological observations. This is why we have chosen $\lambda_{\Psi}^{\text{eff}}$ to be small here.}. 
One can see that $\lambda_{\Psi}^{\text{eff}}$ can cross zero, and its scale strongly depends on $g_{B-L}(\Lambda_{\text{MPP}})$. For convenience, we also show the corresponding effective potentials of $\Psi$ in the upper and middle right panels. Here, we have normalized the vertical axes so that the minimums of the potentials can be easily understood. In the following discussion, besides $\lambda^{\text{eff}}_{\Psi}(\Lambda_{\text{MPP}})>0$ and $\beta_{\lambda^{\text{eff}}_{\Psi}}(\Lambda_{\text{MPP}})>0$, we consider the situation such that only $\lambda^{\text{eff}}$, $\beta_{\lambda}^{\text{eff}}$ and $\kappa$ satisfy the MPP conditions: 
\begin{align} &\lambda^{\text{eff}}(\Lambda_{\text{MPP}})=\beta_{\lambda}^{\text{eff}}(\Lambda_{\text{MPP}})=\kappa(\Lambda_{\text{MPP}})=0\nonumber\\
\lambda^{\text{eff}}_{\Psi}&(\Lambda_{\text{MPP}})>0\h{2mm},\h{2mm}\beta_{\lambda_{\Psi}}^{\text{eff}}(\Lambda_{\text{MPP}})>0\h{2mm},\h{2mm}\beta_{\kappa}(\Lambda_{\text{MPP}})>0.\label{eq:breakingMPP}\end{align}
\\
{\it{\textbf{Question2}}} :  Although we have seen that the B-L symmetry breaking is possible if we break the MPP, is it possible to realize $v_{h}={\cal{O}}(100)$GeV ? 
\\
\\
To answer this question, we should know the typical values of $\kappa$ at a low energy scale (see Eq.(\ref{eq:vev})). Before seeing the numerical results, let us understand it qualitatively. Because we now consider the MPP, the one-loop part of $\beta_{\kappa}$ approximately becomes (see Eq.(\ref{eq:betak}) in \ref{app:beta})
\be \beta_{\kappa}|_{\text{1-loop}}\simeq\frac{12g_{B-L}^{2}g_{\text{mix}}^{2}}{16\pi^{2}}\simeq\frac{g_{B-L}^{2}}{\pi}\times10^{-2},\e
where we have used $g_{\text{mix}}\simeq0.2$ which was obtained from $\lambda^{\text{eff}}(\Lambda_{\text{MPP}})=\beta_{\lambda}^{\text{eff}}(\Lambda_{\text{MPP}})=0$. Thus, $\kappa$ at a low energy scale $\mu$ is approximately given by
\be -\kappa(\mu)=c\times0.1\times g_{B-L}^{2}(\mu),\label{eq:kappaq}\e
where $c$ is a constant and we have used the fact that $g_{B-L}$ does not change significantly. 
\begin{figure}
\begin{center}
\begin{tabular}{c}
\begin{minipage}{0.5\hsize}
\begin{center}
\includegraphics[width=9cm]{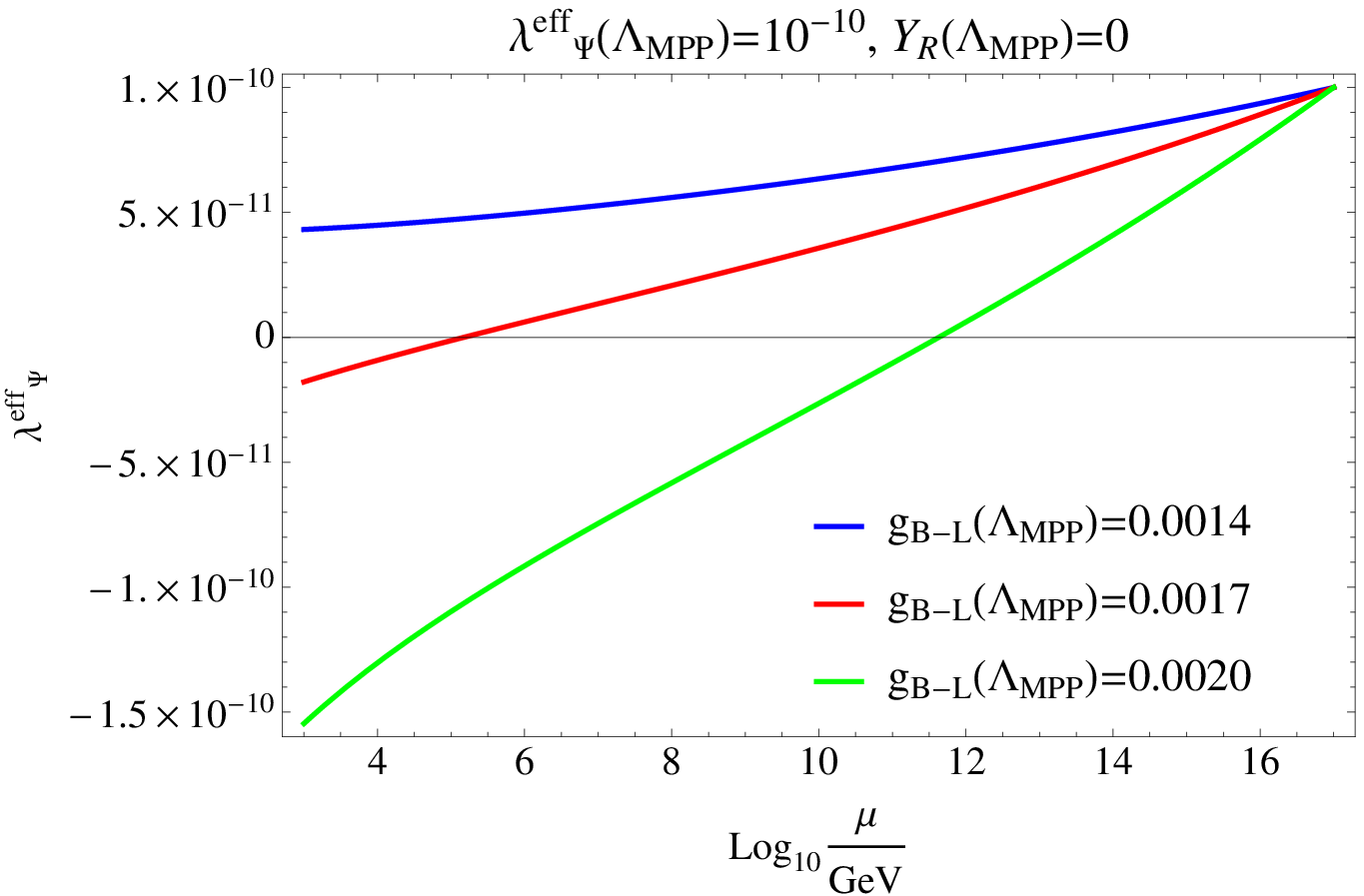}
\end{center}
\end{minipage}
\begin{minipage}{0.5\hsize}
\begin{center}
\includegraphics[width=8.5cm]{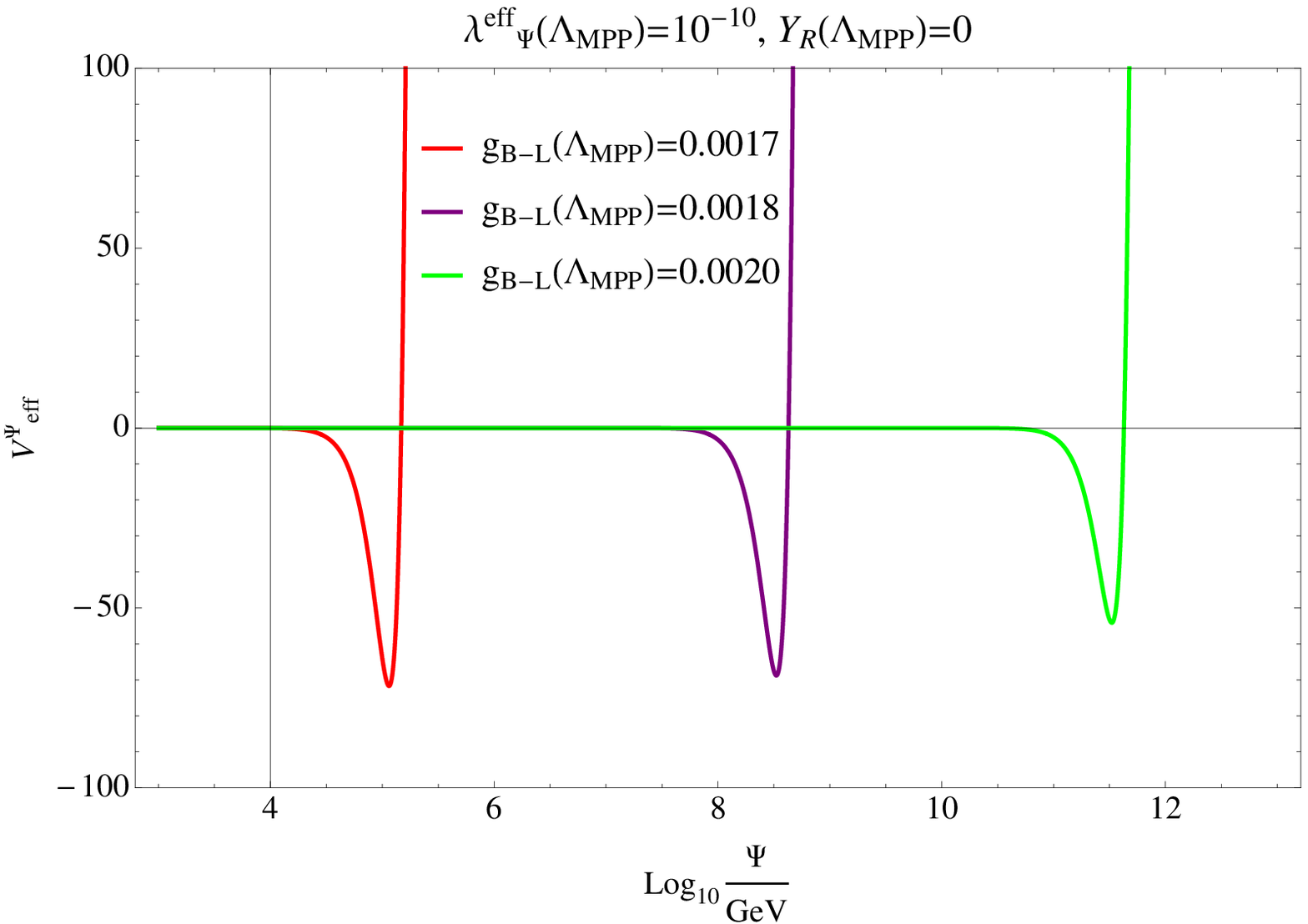}
\end{center}
\end{minipage}
\\
\\
\begin{minipage}{0.5\hsize}
\begin{center}
\includegraphics[width=9cm]{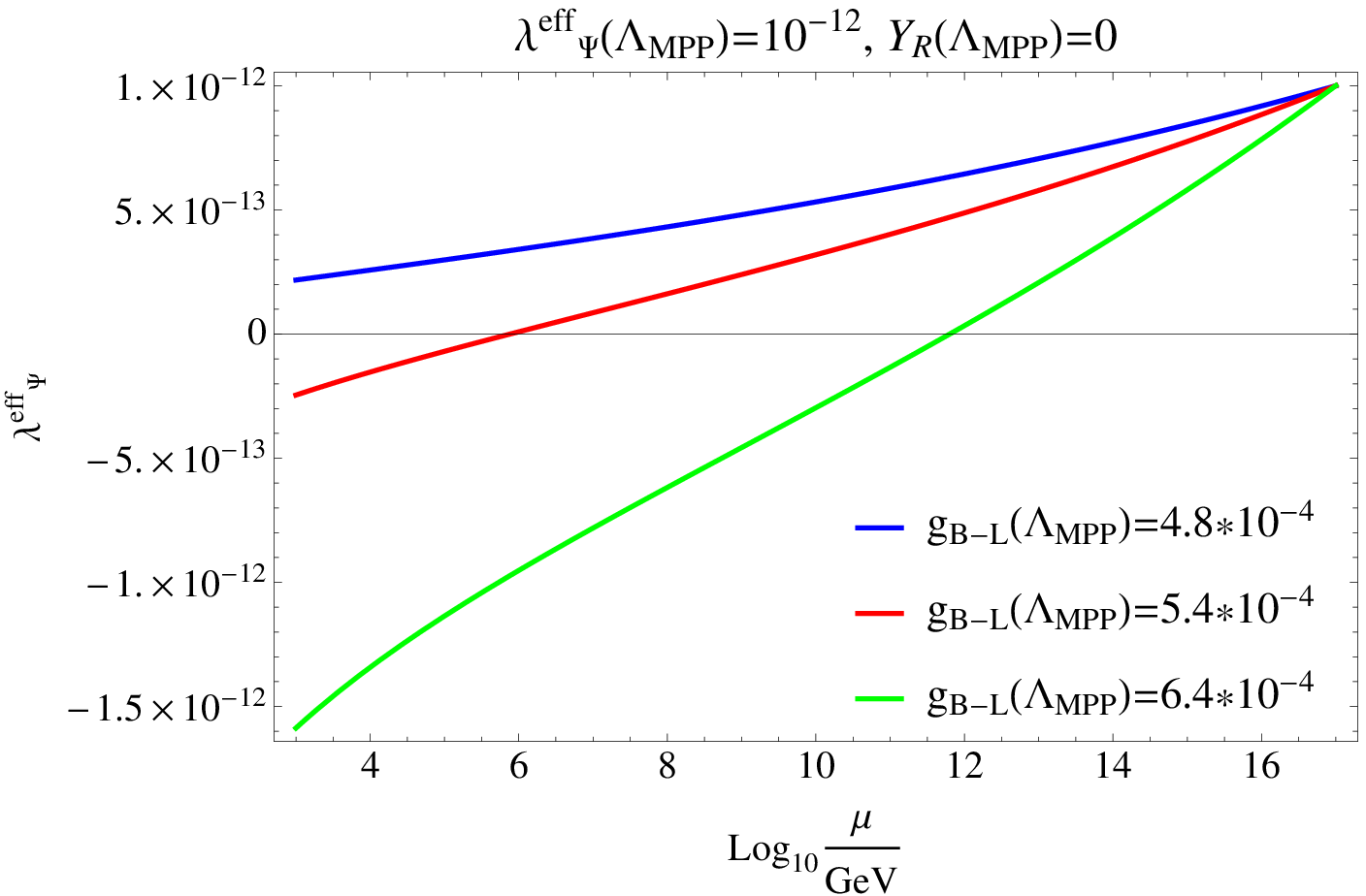}
\end{center}
\end{minipage}
\begin{minipage}{0.5\hsize}
\begin{center}
\includegraphics[width=8.5cm]{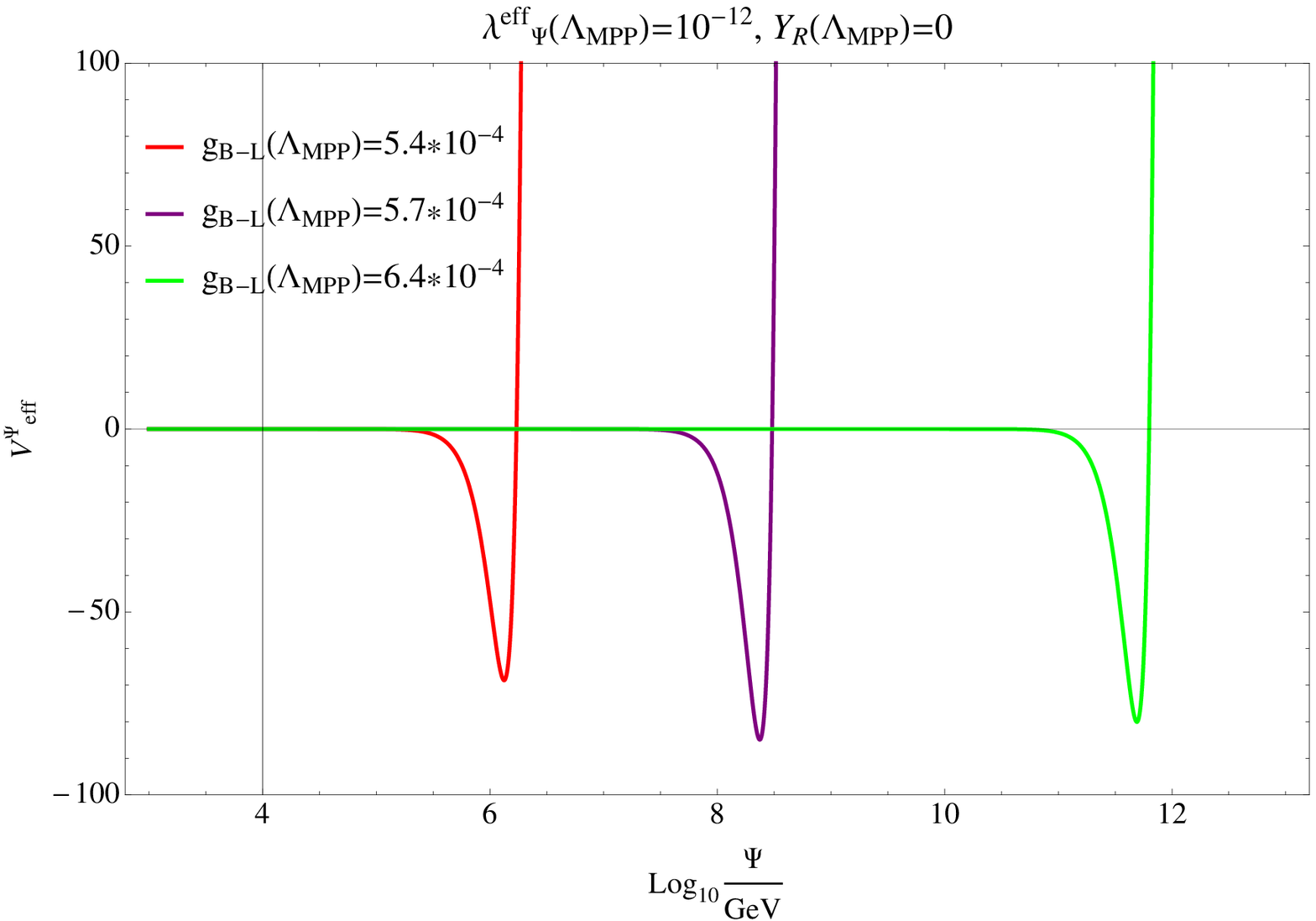}
\end{center}
\end{minipage}
\\
\\ \thispagestyle{empty}
\begin{minipage}{0.5\hsize}
\begin{center}
\includegraphics[width=9cm]{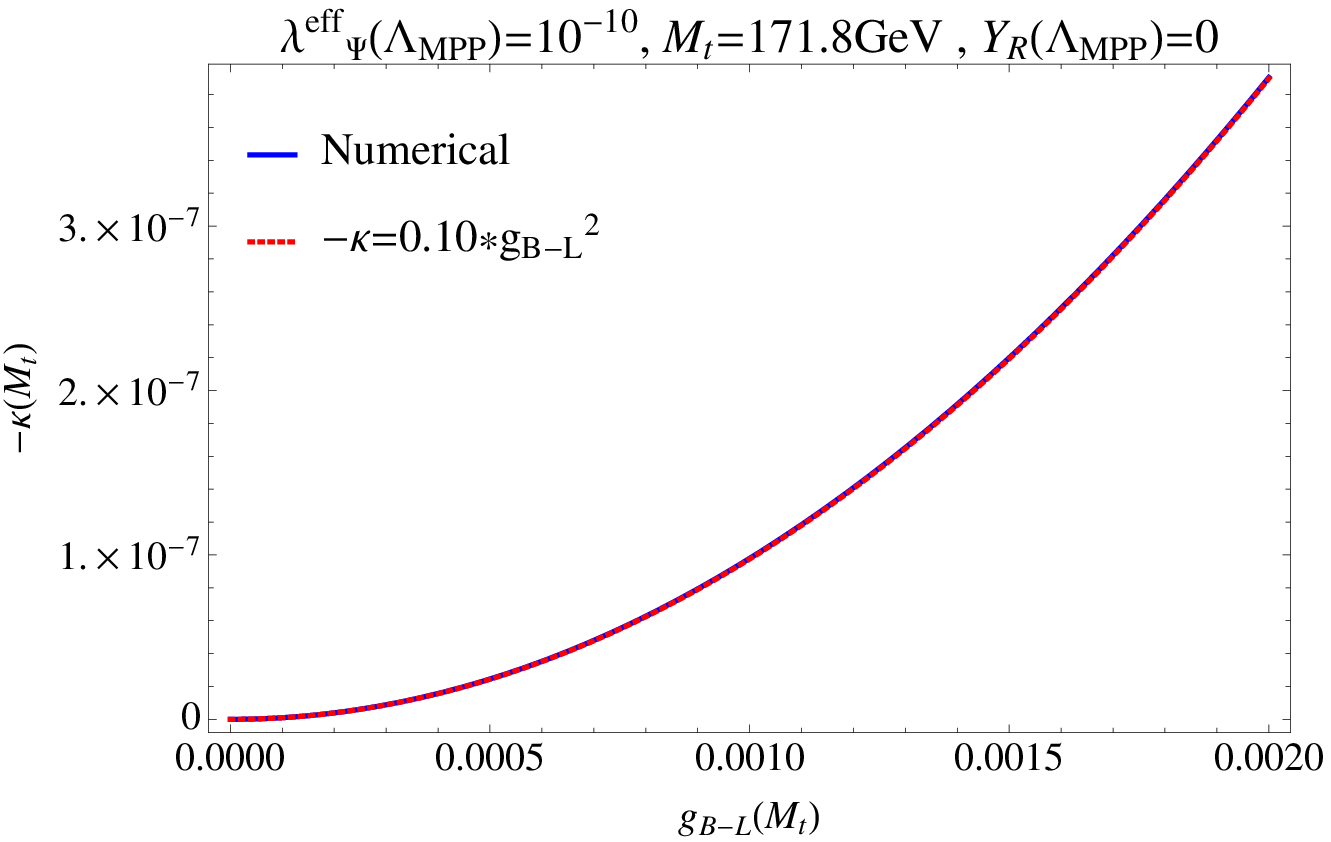}
\end{center}
\end{minipage}
\begin{minipage}{0.5\hsize}
\begin{center}
\includegraphics[width=9cm]{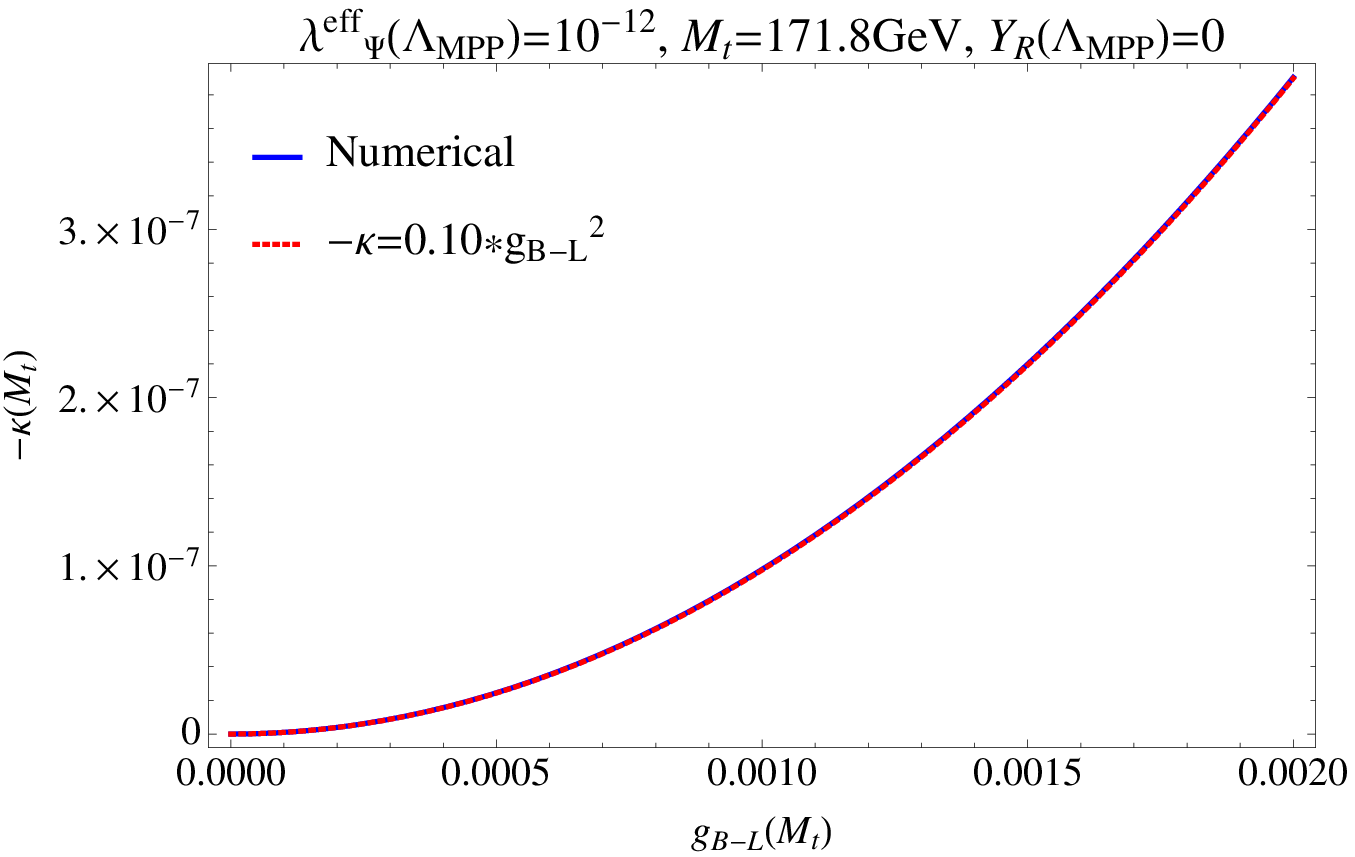}
\end{center}
\end{minipage}
\end{tabular}
\end{center}
\caption{Upper (Middle): the runnings of $\lambda_{\Psi}^{\text{eff}}$ (left) and the corresponding effective potentials (right) in the case of $\lambda_{\Psi}^{\text{eff}}(\Lambda_{\text{MPP}})=10^{-10}\h{1mm}(10^{-12})$. Here, the vertical axes of the right panels are properly normalized. The different colors correspond to the different values of $g_{B-L}(\Lambda_{\text{MPP}})$. Lower left (right): $\kappa$ vs $g_{B-L}$ at $M_{t}=171.8$GeV for $\lambda_{\Psi}^{\text{eff}}(\Lambda_{\text{MPP}})=10^{-10}\h{1mm}(10^{-12})$. Here, the solid blue lines are the numerical results of the RGEs, and the dashed red contours represent $-\kappa=0.10\times g_{B-L}^{2}$.}
\label{fig:Psi}
\end{figure}
\noindent This is the qualitative expression of $\kappa(\mu)$. In the lower left (right) panel of Fig.\ref{fig:Psi}, we show $\kappa$ vs $g_{B-L}$ at $\mu=M_{t}=171.8$GeV in the case of $\lambda_{\Psi}^{\text{eff}}(\Lambda_{\text{MPP}})=10^{-10}\h{1mm}(10^{-12})$. One can see that Eq.(\ref{eq:kappaq}) nicely explains the numerical results when $c$ is $1.0$.
As a result, $v_{h}$ is given by
\be v_{h}=\sqrt{\frac{0.1\times c\times g_{B-L}(v_{h})^{2}}{2\lambda(v_{h})}}\times v_{B-L}\simeq g_{B-L}(v_{h})v_{B-L},\label{eq:vevq}\e
where we have used the typical value $\lambda(v_{h})\simeq0.1$. Therefore, we can obtain $v_{h}={\cal{O}}(100)$GeV by tuning $g_{B-L}(\Lambda_{\text{MPP}})$ and $Y_{R}(\Lambda_{\text{MPP}})$ so that the right hand side of Eq.(\ref{eq:vevq}) becomes ${\cal{O}}(100)$GeV. The red lines of the upper and middle left panels of Fig.\ref{fig:Psi} show such examples. In the upper (lower) case, $g_{B-L}$ is ${\cal{O}}(10^{-3}(10^{-4}))$ and $v_{B-L}$ is ${\cal{O}}(10^{2}(10^{3}))$TeV.

\noindent

\begin{figure}
\begin{center}
\begin{tabular}{c}
\begin{minipage}{0.5\hsize}
\begin{center}
\includegraphics[width=8.5cm]{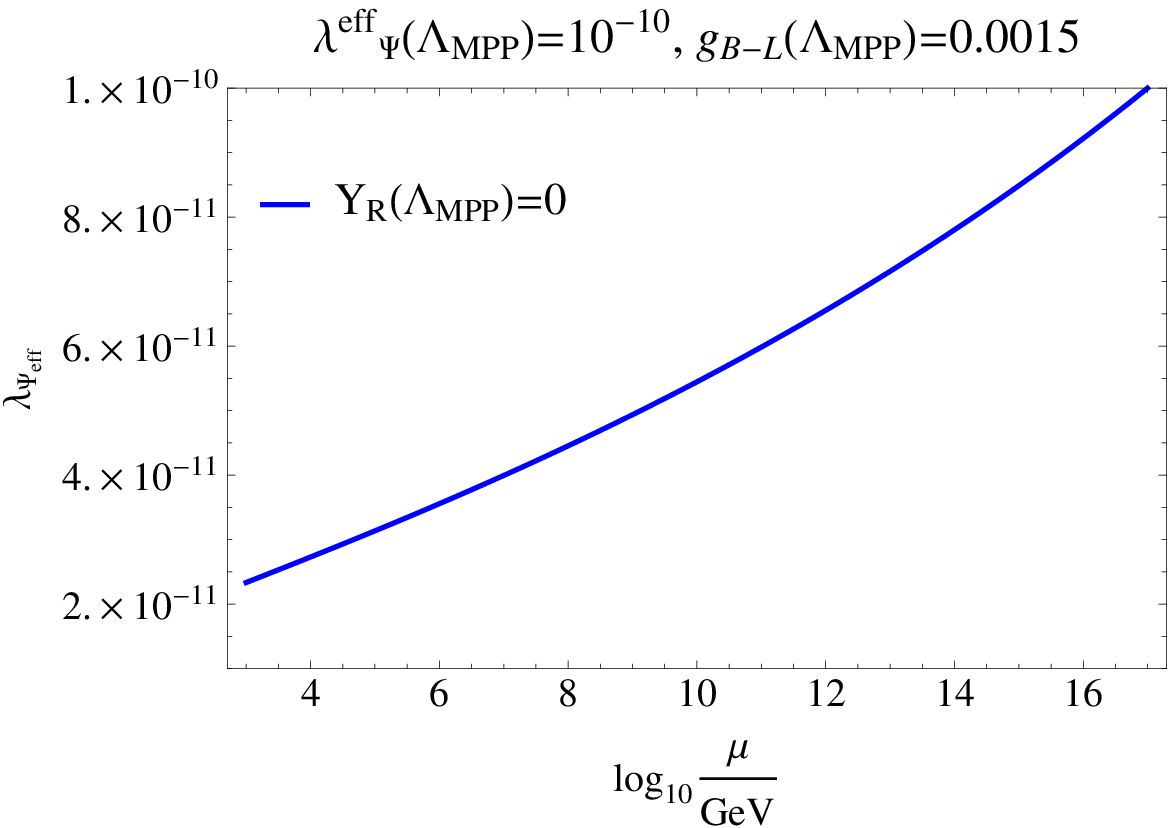}
\end{center}
\end{minipage}
\begin{minipage}{0.5\hsize}
\begin{center}
\includegraphics[width=8.5cm]{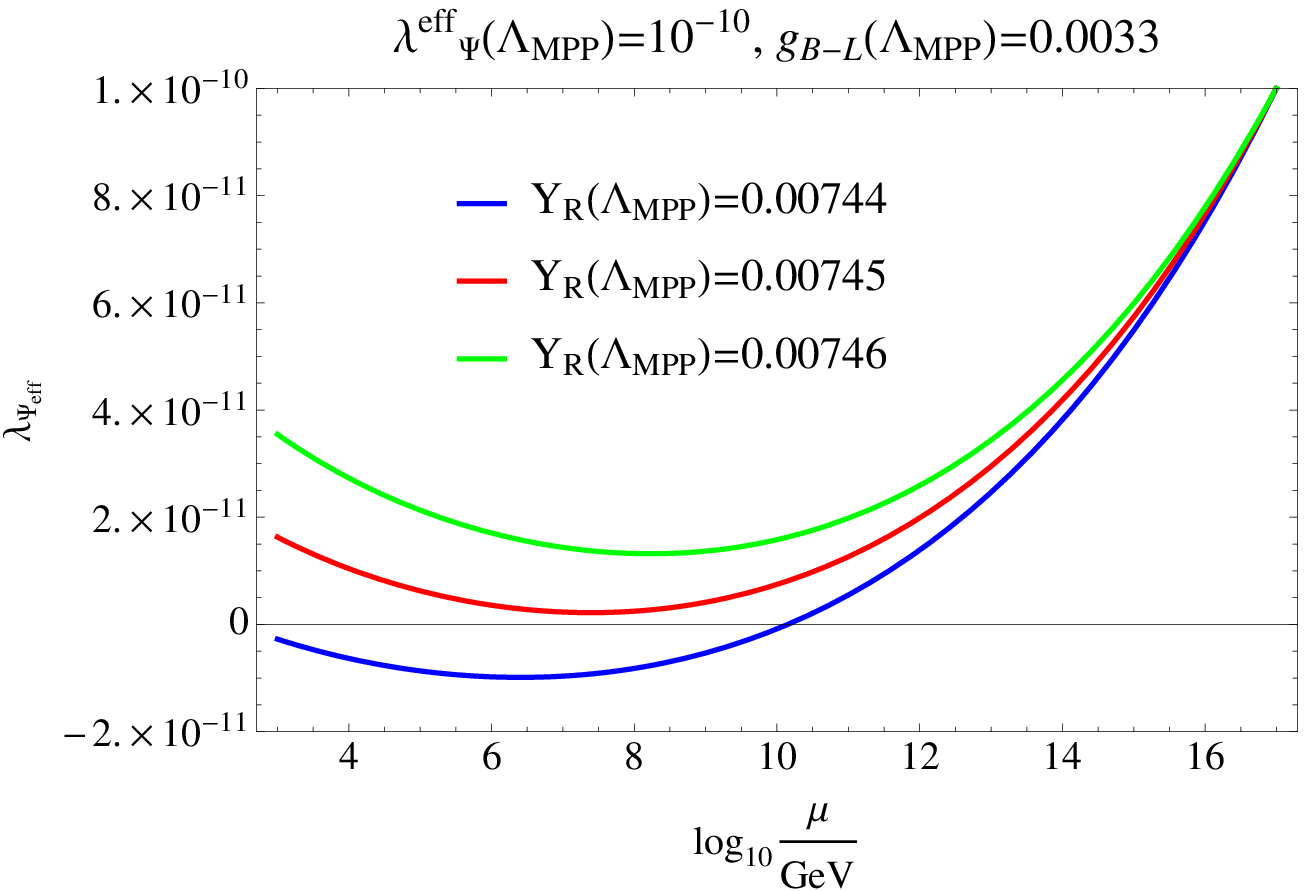}
\end{center}
\end{minipage}
\end{tabular}
\end{center}
\caption{The impossibility to realize $v_{h}={\cal{O}}(100)$GeV when $g_{B-L}(\Lambda_{\text{MPP}})$ is outside the region given by Eq.(\ref{eq:gblre}). The left (right) panel shows the running of $\lambda_{\Psi}^{\text{eff}}$ when $g_{B-L}(\Lambda_{\text{MPP}})=0.0015\h{1mm}(0.0033)$. In the left panel, one can see that $\lambda_{\Psi}^{\text{eff}}$ is always positive even if $Y_{R}(\Lambda_{\text{MPP}})=0$. In the right panel, one can see that B-L symmetry breaking occurs at a very high energy scale ($\gg10^{2}$ TeV).}
\label{fig:const}
\end{figure}

A few comments are needed. First, because we no longer impose the flatness of $V^{\Psi}_{\text{eff}}$, the two parameters $g_{B-L}(\Lambda_{\text{MPP}})$ and $Y_{R}(\Lambda_{\text{MPP}})$ are remaining as free parameters. However, the parameter region which can produce $v_{h}={\cal{O}}(100)$GeV is quite limited. For example, in the $\lambda^{\text{eff}}_{\Psi}(\Lambda_{\text{MPP}})=10^{-10}$ case, it is 
\be 1.6\times10^{-3}\lesssim g_{B-L}(\Lambda_{\text{MPP}})\lesssim 3.2\times10^{-3},\label{eq:gblre}\e
and $Y_{R}(\Lambda_{\text{MPP}})$ is correspondingly fixed so that $\lambda_{\Psi}^{\text{eff}}$ crosses zero around ${\cal{O}}(100)$TeV. The reason for this is as follows. When $g_{B-L}(\Lambda_{\text{MPP}})$ is small, $\beta_{\lambda_{\Psi}}$ is too small to make $\lambda^{\text{eff}}_{\Psi}$ negative at a low energy scale. As a result, the B-L symmetry breaking does not occur. On the other hand, when $g_{B-L}(\Lambda_{\text{MPP}})$ is too large, the B-L symmetry breaking occurs at a very high energy scale. We can actually see these behaviors from Fig.\ref{fig:const}. Note that the allowed values of $g_{B-L}(\Lambda_{\text{MPP}})$ become small when we decrease $\lambda_{\Psi}^{\text{eff}}(\Lambda_{\text{MPP}})$.

\begin{figure}
\begin{center}
\includegraphics[width=9cm]{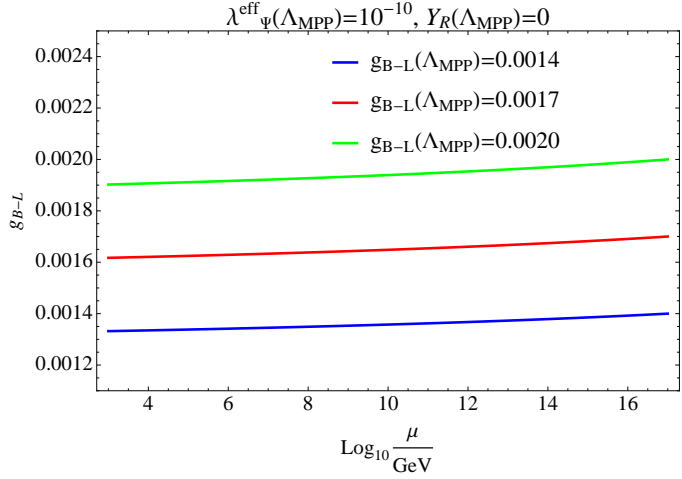}
\end{center}
\caption{The typical runnings of $g_{B-L}$ when $\lambda_{\Psi}^{\text{eff}}(\Lambda_{\text{MPP}})=10^{-10}$. 
}
\label{fig:typicalgbl}
\end{figure}

Second, $g_{B-L}$ at a low energy scale does not change very much from the value at $\Lambda_{\text{MPP}}$. See Fig.\ref{fig:typicalgbl}. This shows the typical runnings of $g_{B-L}$ when $\lambda_{\Psi}^{\text{eff}}(\Lambda_{\text{MPP}})=10^{-10}$. 

Finally, when Eq.(\ref{eq:breakingMPP}) is satisfied, the mass of the B-L gauge boson is uniquely predicted to be 
\be M_{B-L}=2g_{B-L}(v_{B-L})v_{B-L}=2\sqrt{2}\times\sqrt{\frac{\lambda(v_{h})}{0.10}}\times v_{h},\label{eq:masspre}\e 
where we have used Eq.(\ref{eq:vevq}) and $c=1.0$. By using the experimental value $v_{h}=246$GeV and the typical value $\lambda(v_{h})\simeq0.1$, this leads to
\be M_{B-L}\simeq 696\h{1mm}\text{GeV}.\e
Although this is a remarkable prediction of the MPP, this value is already excluded by the experiment of ATLAS \cite{Aad:2014cka} because $g_{\text{mix}}$ is too large\footnote{In \cite{Aad:2014cka}, $g_{\text{mix}}$ is represented by $\tilde{g}_{Y}$. Therefore, $g_{\text{mix}}\simeq0.24$ corresponds to the contour $\gamma'\simeq0.32/\sin\theta$ in FIG.7 of \cite{Aad:2014cka}.}. 

\section{Non-Minimal Inflation - SM Singlet Scalar \\ \hspace{10cm}as the Inflaton -}\label{sec:inf}
As is well known, Higgs inflation is possible in the SM \cite{Bezrukov:2007ep,Hamada:2013mya,Hamada:2014iga,Hamada:2014wna,Hamada:2014raa}. There, the criticality of the Higgs potential plays a crucial role to realize the inflation naturally; we can obtain a sufficient e-foldings and the CMB fluctuations even if $\xi$ is ${\cal{O}}(1)$ by making the running Higgs self coupling arbitrary small (see \cite{Hamada:2014wna} for more details). In other words, the smallness of the self coupling is needed to realize the inflation naturally. Such a Higgs inflation is of course possible in our B-L model, however, the conclusion of the previous section indicates a new possibility: The newly introduced SM singlet complex scalar $\Psi$ plays a roll of the inflaton \cite{Okada:2011en}. 
We study this scenario in this section. 

The action with the non-minimal gravitational coupling $\xi\Psi^{2}{\cal{R}}$ in the Jordan frame is given by
\be S_{J}=\int d^{4}x\sqrt{-g}\Biggl\{-\left(\frac{M_{pl}^{2}+\xi \Psi^{2}}{2}\right){\cal{R}}+\frac{1}{2}g^{\mu\nu}\partial_{\mu}\Psi\partial_{\nu}\Psi-\frac{\lambda^{\text{eff}}_{\Psi}(\Psi)}{4}\Psi^{4}+\cdots\Biggl\},\e
where $\Psi$ is the physical (real) field, and we have written the relevant terms for the later discussion. To study the inflation, it is convenient to move to the Einstein frame. Namely, by the conformal transformation 
\be g_{\mu\nu}^{\text{E}}:=\Omega^{2}g_{\mu\nu}\h{2mm},\h{2mm}\Omega^{2}:=1+\frac{\xi\Psi^{2}}{M_{pl}^{2}},\e
and the field redefinition 
\be \frac{d\chi}{d\Psi}=\sqrt{\frac{\Omega^{2}+6\xi^{2}\Psi^{2}/M_{pl}^{2}}{\Omega^{4}}},\label{eq:chi}\e
the action becomes
\be S_{E}=\int d^{4}x\sqrt{-g_{E}}\Biggl\{-\frac{M_{pl}^{2}}{2}{\cal{R}}_{E}+\frac{1}{2}g_{E}^{\mu\nu}\partial_{\mu}\chi\partial_{\nu}\chi-\frac{\lambda^{\text{eff}}_{\Psi}(\Psi)}{4\Omega^{4}}\Psi^{4}(\chi)+\cdots\Biggl\}.\e

\begin{figure}
\begin{center}
\begin{tabular}{c}
\begin{minipage}{0.5\hsize}
\begin{center}
\includegraphics[width=8.5cm]{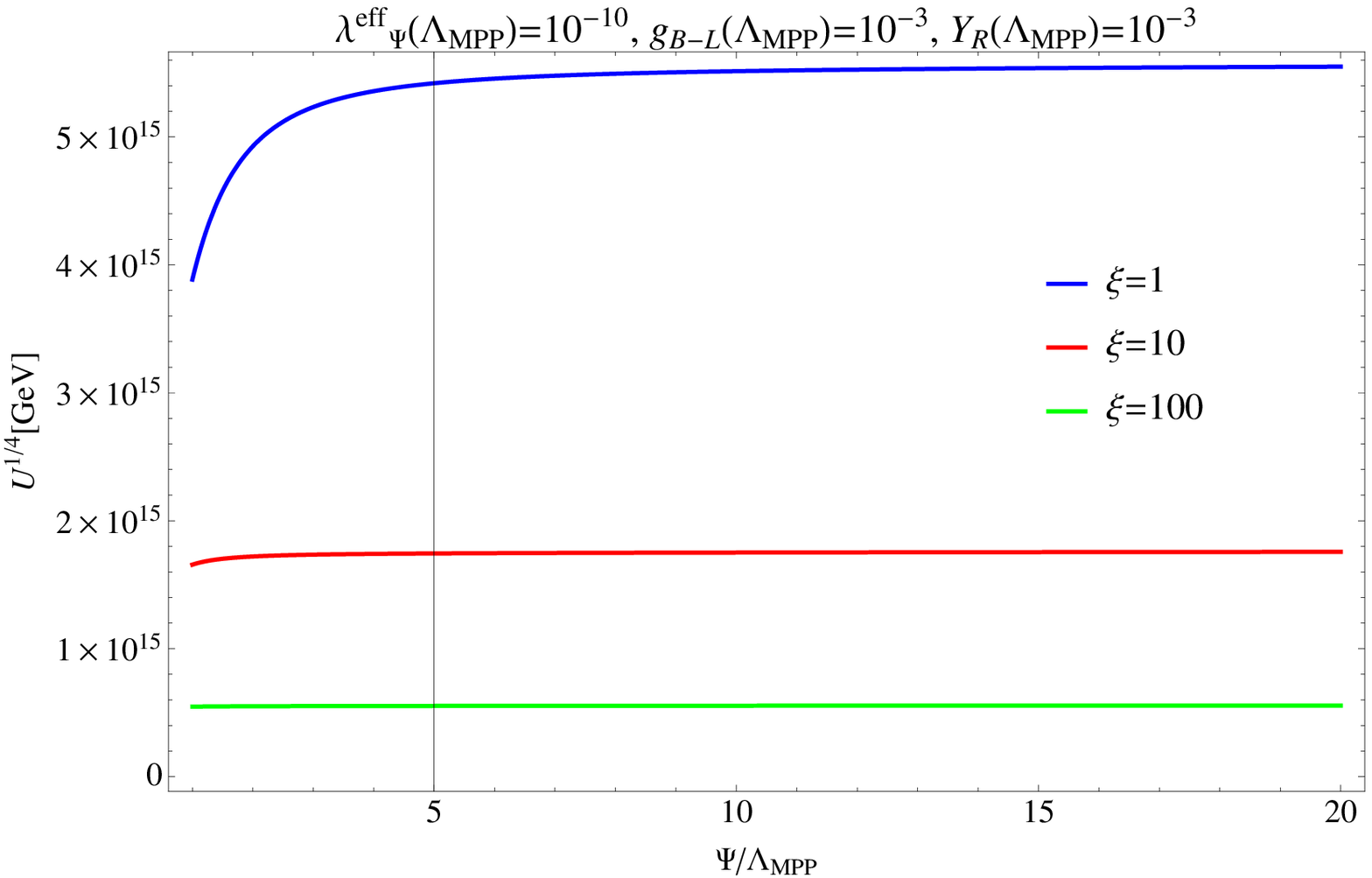}
\end{center}
\end{minipage}
\begin{minipage}{0.5\hsize}
\begin{center}
\includegraphics[width=8.5cm]{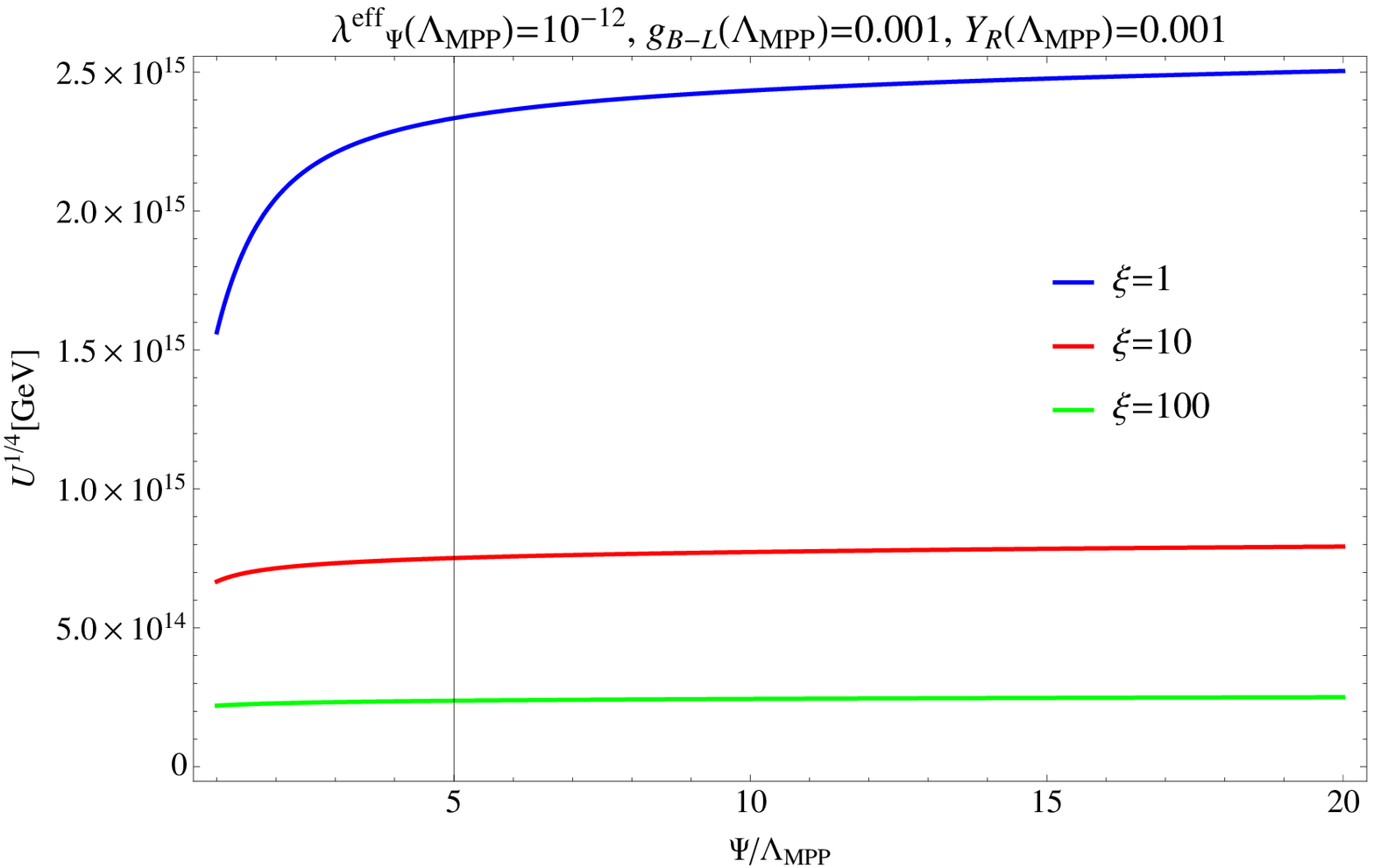}
\end{center}
\end{minipage}
\end{tabular}
\end{center}
\caption{The effective potentials in the Einstein frame. The left (right) panel shows the $\lambda_{\Psi}^{\text{eff}}(\Lambda_{\text{MPP}})=10^{-10}\h{1mm}(10^{-12})$ case. The different colors correspond to the different values of $\xi$.
}
\label{fig:potein}
\end{figure}

\noindent This is the canonically normalized form, and the potential in this frame is given by
\be U(\chi):=\frac{\lambda^{\text{eff}}_{\Psi}(\Psi)}{4\Omega^{4}}\Psi^{4}(\chi).\label{eq:pot1}\e
For the large values of $\Psi\gg M_{pl}/\sqrt{\xi}$, Eq.(\ref{eq:chi}) becomes
\be \frac{d\chi}{d\Psi}\simeq \frac{M_{pl}}{\Psi}\sqrt{\frac{1+6\xi}{\xi}},\e
so we have
\be \Psi\simeq M_{pl}\exp\left(\frac{\chi}{M_{pl}\sqrt{(1+6\xi)/\xi}}\right).\e
In this limit, the potential in the Einstein frame Eq.(\ref{eq:pot1}) becomes
\be U(\chi)\simeq \frac{\lambda^{\text{eff}}_{\Psi}(\Psi)M_{pl}^{4}}{4\xi^{2}}\left(1+\exp\left(-\frac{2\chi}{M_{pl}\sqrt{(1+6\xi)/\xi}}\right)\right)^{-2}.\e
This is an exponentially flat potential (see Fig.(\ref{fig:potein}) for example), so we can use the slow-roll approximations. The slow-roll parameters are
\begin{align} &\epsilon:=\frac{M_{pl}^{2}}{2}\left(\frac{1}{U}\frac{dU}{d\chi}\right)^{2}=\frac{M_{pl}^{2}}{2}\left(\frac{d\Psi}{d\chi}\frac{U'}{U}\right)^{2},\\
&\eta:=M_{pl}^{2}\left(\frac{1}{U}\frac{d^{2}U}{d\chi^{2}}\right)=\frac{M_{pl}^{2}}{U}\frac{d\Psi}{d\chi}\frac{d}{d\Psi}\left(\frac{d\Psi}{d\chi}U'\right),\\
&\zeta^{2}:=M_{pl}^{4}\frac{1}{U^{2}}\frac{d^{3}U}{d\chi^{3}}\frac{dU}{d\chi},\end{align}
where a prime represents a derivative with respect to $\Psi$. By using these quantities, the number of e-foldings $N$, the spectral index $n_{s}$, its running $dn_{s}/d\ln k$ and the tensor-to-scalar ratio $r$ are given by
\begin{align} N&=\int_{\chi_{end}}^{\chi_{ini}}d\chi\frac{1}{M_{pl}^{2}}\frac{U}{dU/d\chi}=\int_{\Psi_{end}}^{\Psi_{ini}}d\Psi\frac{1}{M_{pl}^{2}}\frac{U}{U'}\\
n_{s}&=1-6\epsilon+2\eta,\\
\frac{dn_{s}}{d\ln k}&=16\epsilon\eta-24\epsilon^{2}-2\zeta^{2},\\
r&=16\epsilon,\end{align}
where $\Psi_{ini}\h{1mm}(\Psi_{end})$ represents the initial (end) value of $\Psi$. In the following discussion, we denote $\Psi_{ini}$ simply as $\Psi$.

Here, we give the current cosmological constraints by Planck TT + lowP \cite{Planck:2015xua}. The overall normalization of the CMB fluctuations at the scale $k_{0}=0.05$ Mpc$^{-1}$ is 
\be A_{s}:=\frac{U}{24\pi^{2}\epsilon M_{pl}^{4}}\Biggl|_{k_{0}}=\left(2.198^{+0.076}_{-0.085}\right)\times10^{-9}\h{2mm}(\text{68\% CL}),\e
and $n_{s}$, $dn_{s}/d\ln k$ and $r$ are
\be n_{s}=0.9655\pm 0.0062\h{1mm}(\text{68\% CL}),\h{2mm}\frac{dn_{s}}{d\ln k}=-0.0126^{+0.0098}_{-0.0087}\h{1mm}(\text{68\% CL}),\h{2mm}r_{0.002}<0.10\h{1mm}(\text{95\% CL}),\label{eq:planck}\e
at the scale $k_{0}=0.05$ Mpc$^{-1}$ for $n_{s}$ and $dn_{s}/d\ln k$, and $k_{r}=0.002$ Mpc$^{-1}$ for $r_{0.002}$. On the other hand, the BICEP2 experiment has reported an observation of $r_{0.002}$ \cite{Ade:2014xna}:
\be r_{0.002}=0.20^{+0.07}_{-0.05} \h{2mm}(\text{68\% CL}).\label{eq:bicep}\e
There has been discussion such that this result may be consistent with $r=0$ due to the foreground effect \cite{Mortonson:2014bja,Flauger:2014qra}.
\\

Our calculations are based on the following conditions:
\begin{enumerate}\item Although there are six parameters, we consider the situation where Eq.(\ref{eq:breakingMPP}) is satisfied. Namely, $M_{t}$,   $g_{\text{mix}}(\Lambda_{\text{MPP}})$ and $\kappa(\Lambda_{\text{MPP}})$ are fixed respectively at $171.8$GeV, $0.2$ and $0$. 
\item As the typical values of $\lambda_{\Psi}^{\text{eff}}(\Lambda_{\text{MPP}})$, we choose
\be \lambda_{\Psi}^{\text{eff}}(\Lambda_{\text{MPP}})=10^{-10},\h{1mm}10^{-12}\h{1mm}\text{and}\h{1mm}10^{-14}.\e
\item The remaining two parameters $g_{B-L}(\Lambda_{\text{MPP}})$ and $Y_{R}(\Lambda_{\text{MPP}})$ are chosen so that $v_{h}$ becomes ${\cal{O}}(100)$GeV. As discussed at the end of Section \ref{sec:b-l}, the allowed region is quite limited in this case. We have checked that the cosmological predictions do not change very much even if we change these parameters within such region(see Fig.\ref{fig:gbl}).\\
\end{enumerate}

\begin{figure}
\begin{center}
\begin{tabular}{c}
\begin{minipage}{0.5\hsize}
\begin{center}
\includegraphics[width=7.5cm]{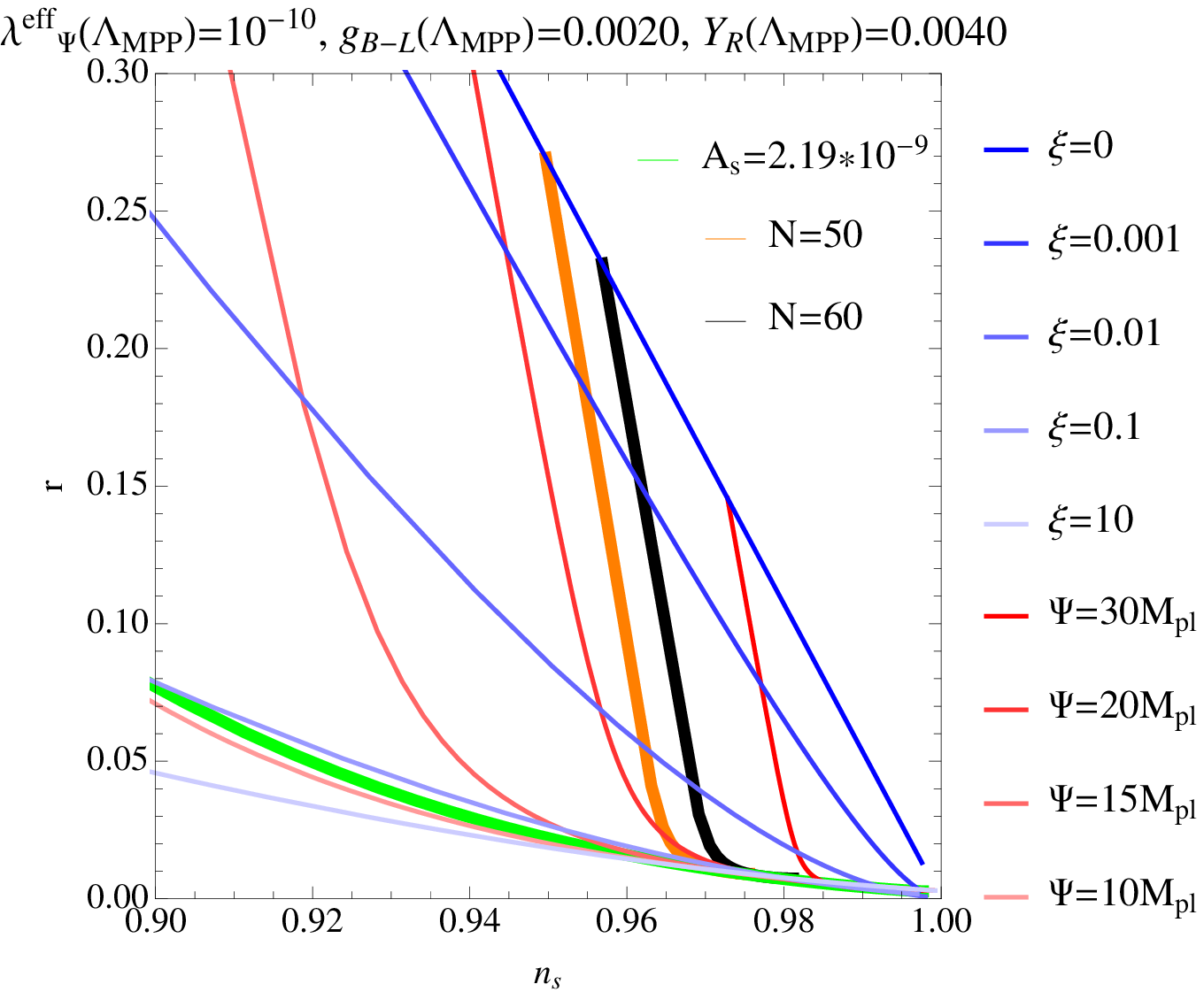}
\end{center}
\end{minipage}
\begin{minipage}{0.5\hsize}
\begin{center}
\includegraphics[width=7.5cm]{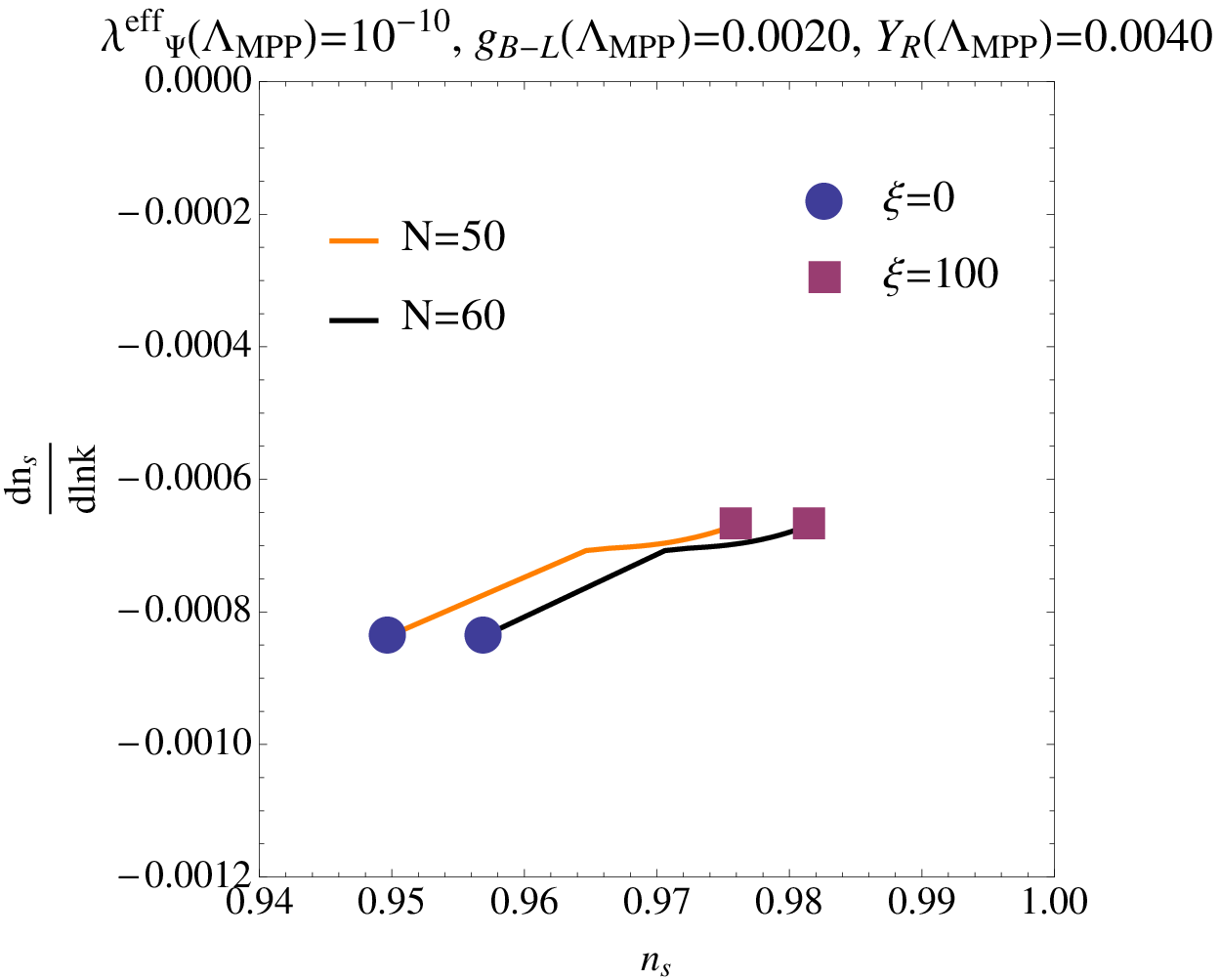}
\end{center}
\end{minipage}
\\
\\
\begin{minipage}{0.5\hsize}
\begin{center}
\includegraphics[width=7.5cm]{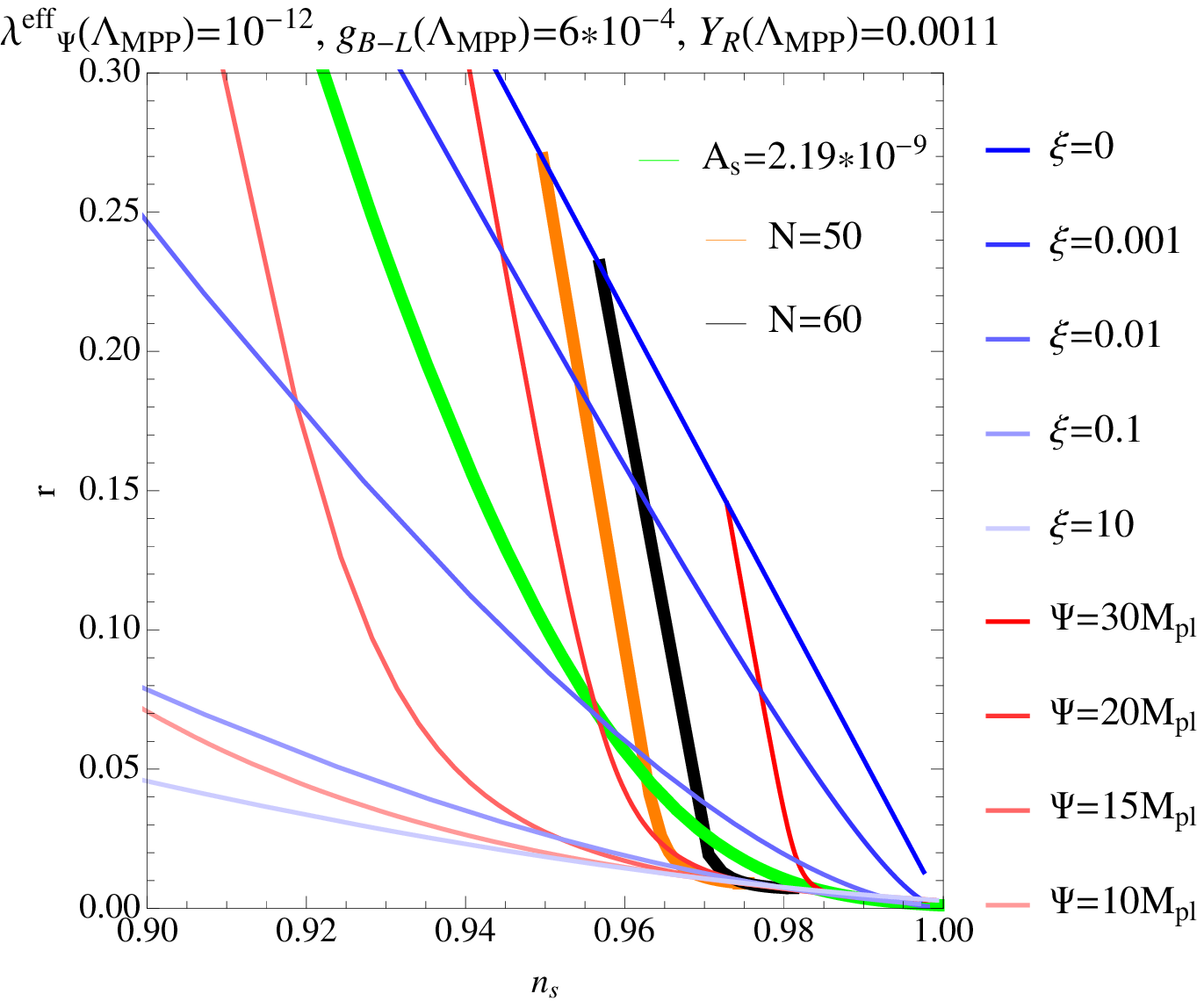}
\end{center}
\end{minipage}
\begin{minipage}{0.5\hsize}
\begin{center}
\includegraphics[width=7.5cm]{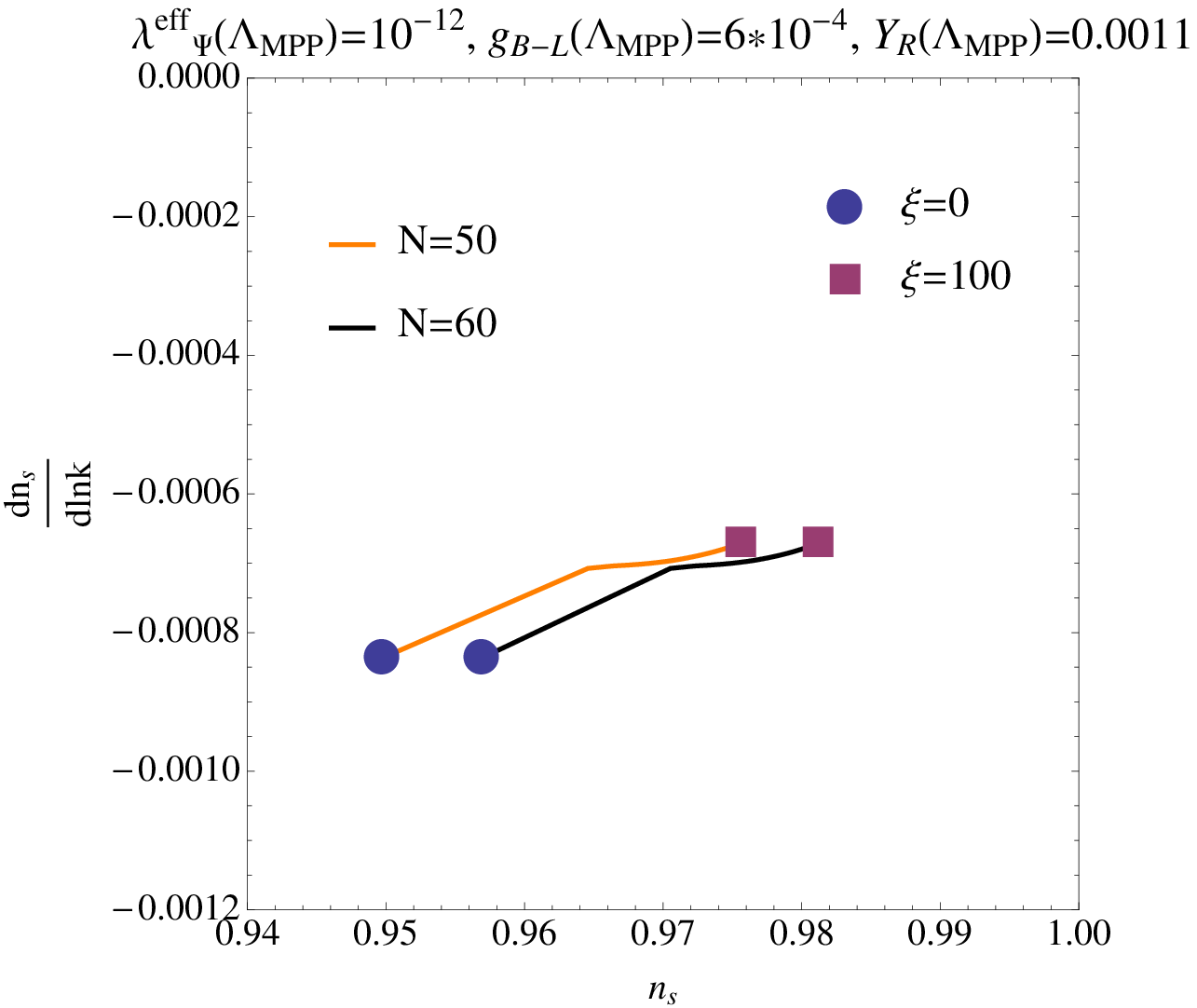}
\end{center}
\end{minipage}
\\
\\
\begin{minipage}{0.5\hsize}
\begin{center}
\includegraphics[width=7.5cm]{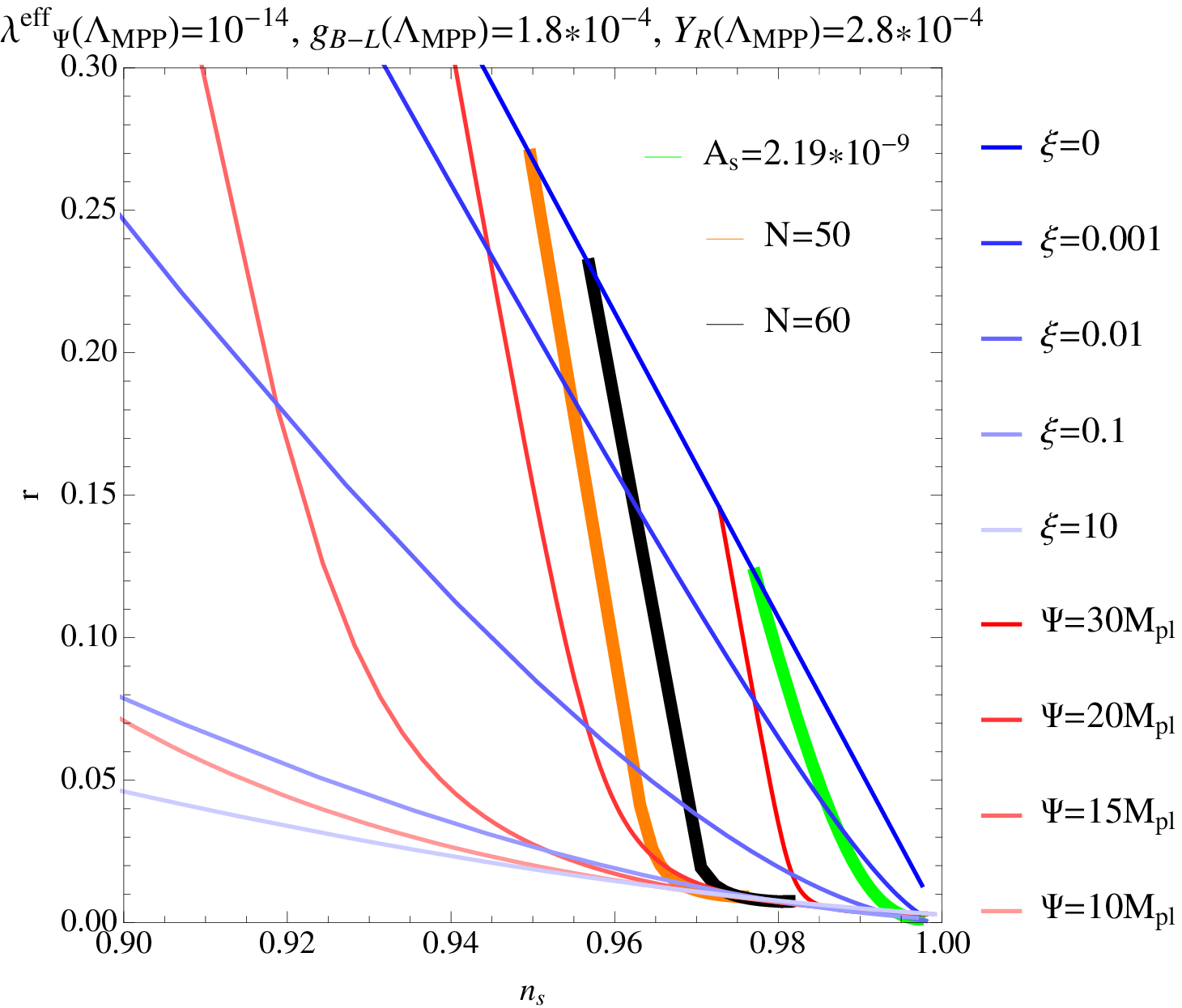}
\end{center}
\end{minipage}
\begin{minipage}{0.5\hsize}
\begin{center}
\includegraphics[width=7cm]{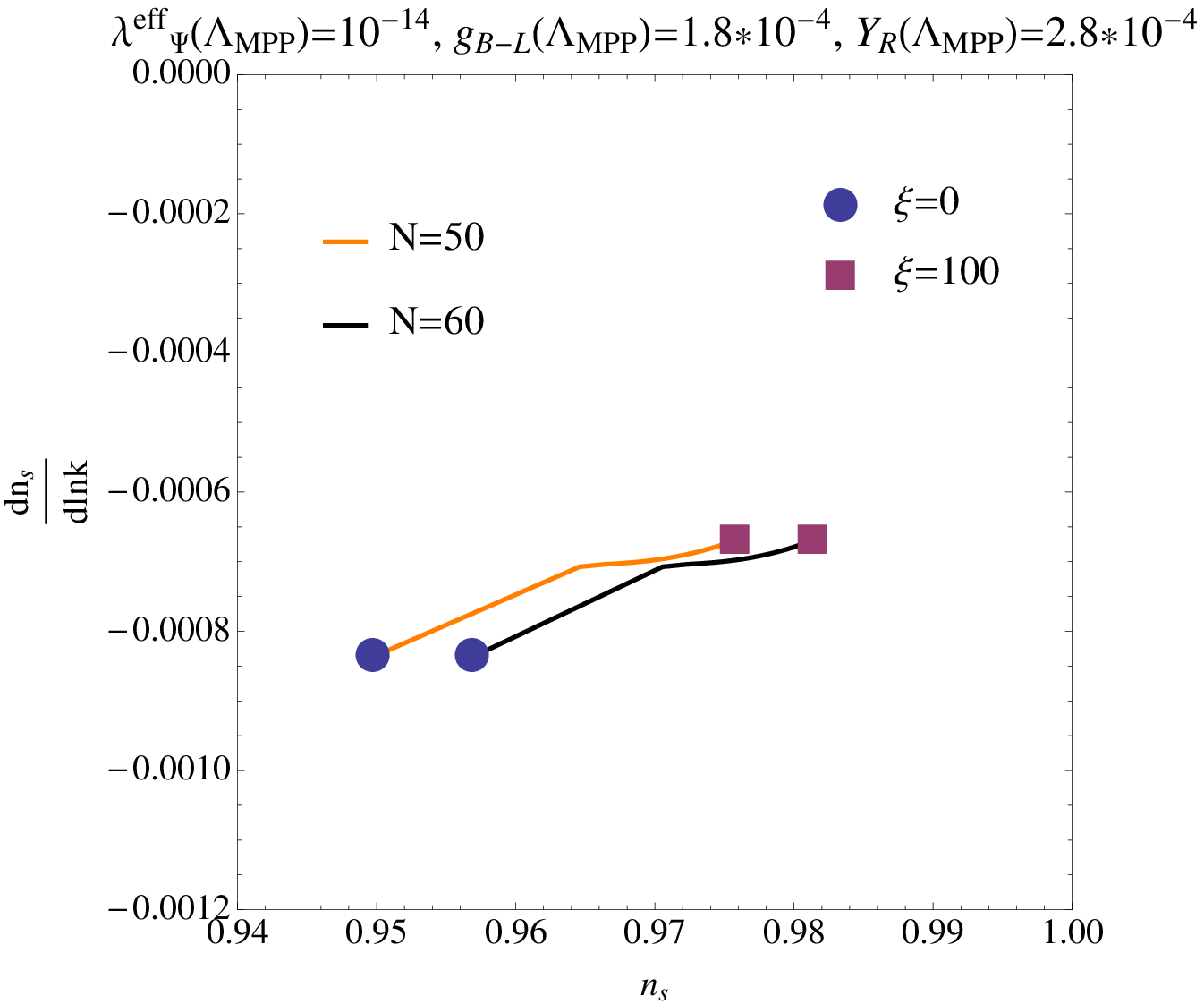}
\end{center}
\end{minipage}
\end{tabular}
\end{center}
\caption{The cosmological predictions of the gauged B-L model. The upper, middle and lower panels correspond to $\lambda_{\Psi}^{\text{eff}}(\Lambda_{\text{MPP}})=10^{-10}$, $10^{-12}$ and $10^{-14}$ respectively. The left (right) panels show $n_{s}$ vs $r\h{1mm}(dn_{s}/\ln k)$. The blue (red) lines indicate $\xi\h{1mm}(\Psi)=constant$, and the contours which correspond to $N=50$ and $60$ are represented by orange and black respectively.}
\label{fig:result}
\end{figure}

\noindent Fig.\ref{fig:result} shows our numerical results when we fix $g_{B-L}(\Lambda_{\text{MPP}})$ and $Y_{R}(\Lambda_{\text{MPP}})$. Our results are, of course, consistent with the previous results such as \cite{Okada:2011en,Okada:2014lxa}. The left (right) panels show $r\h{1mm}(dn_{s}/\ln k)$ vs $n_{s}$. Here, the solid blue (red) lines represent $\xi\h{1mm}(\Psi)=$constant, and the contours which correspond to $N=50$ and $60$ are represented by orange and black respectively from $\xi=0$ to $\xi=100$. In the left panels, we also show the contours of $A_{s}=2.2\times10^{-9}$ by green. These results are consistent with the observed results (\ref{eq:planck})(\ref{eq:bicep}) by Planck and BICEP2. Especially, as one can see from the behaviors of the green lines, the values of $\lambda_{\Psi}^{\text{eff}}(\Lambda_{\text{MPP}})$ which can simultaneously explain $A_{s}=2.19\times10^{-9}$, the sufficient e-foldings ($N\geq 50$) and the BICEP2's result $r=0.2$ are quite limited:
\be 10^{-14}<\lambda_{\Psi}^{\text{eff}}(\Lambda_{\text{MPP}})<10^{-12}.\e
Among the three quantities $n_{s}$, $r$ and $dn_{s}/\ln k$, one might think that the predicted values of $dn_{s}/\ln k$ are small compared with the observed values ${\cal{O}}(-0.01)$. It might be possible to improve this situation by including a higher dimensional operator. See \cite{Hamada:2014wna} for example.

\begin{figure}
\begin{center}
\begin{tabular}{c}
\begin{minipage}{0.5\hsize}
\begin{center}
\includegraphics[width=6.5cm]{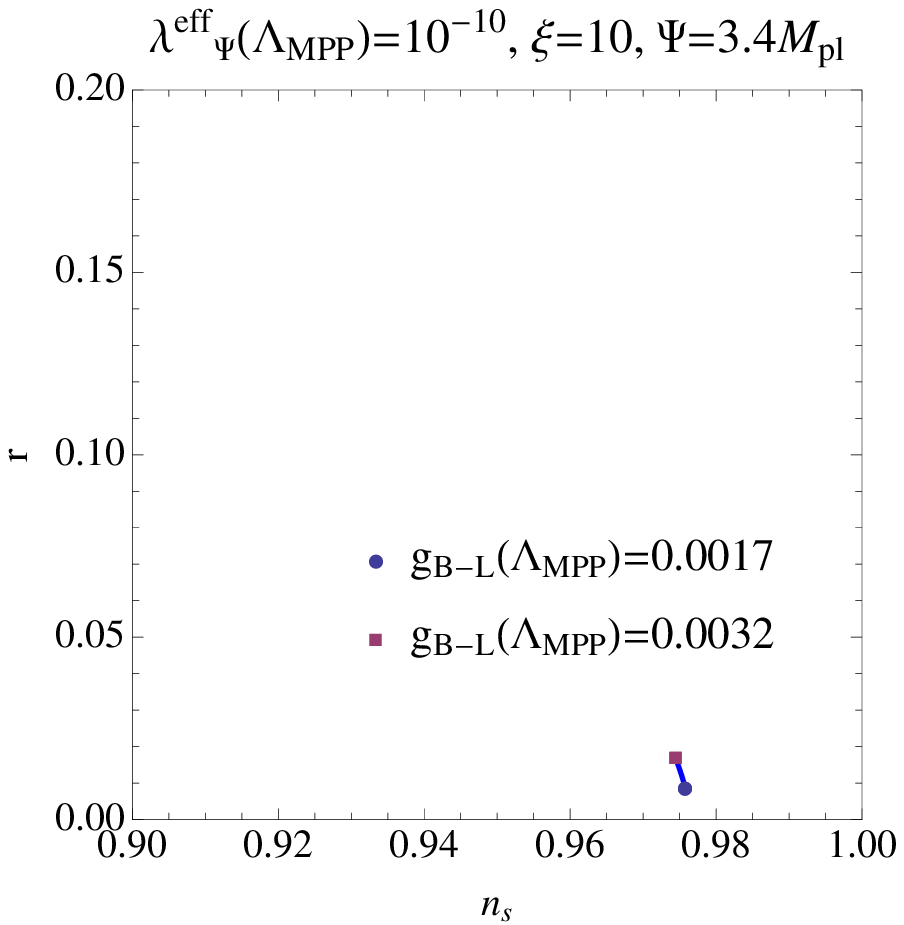}
\end{center}
\end{minipage}
\begin{minipage}{0.5\hsize}
\begin{center}
\includegraphics[width=7cm]{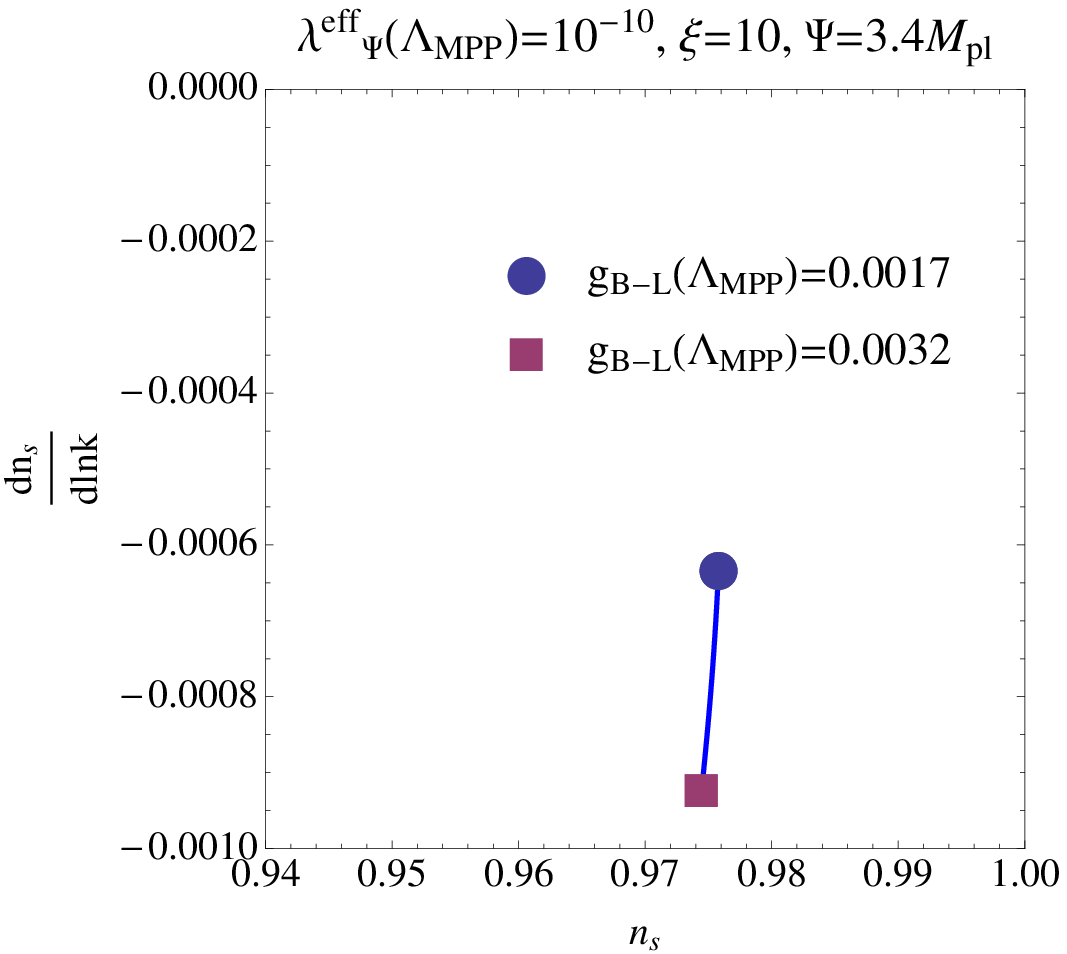}
\end{center}
\end{minipage}
\end{tabular}
\end{center}
\caption{The $g_{B-L}(\Lambda_{\text{MPP}})$ dependences of $n_{s}$, $r$ and $dn_{s/\ln k}$. Here, we change $g_{B-L}(\Lambda_{\text{MPP}})$ within the region such that the electroweak symmetry breaking occurs at ${\cal{O}}(100)$GeV, and $\xi$ and $\Psi$ are chosen so that both of the observed value of $A_{s}$ and $N=50$ are satisfied when $g_{B-L}(\Lambda_{\text{MPP}})=0.0020$. The left (right) panel shows $r\h{1mm}(dn_{s}/\ln k)$ vs $n_{s}$. 
}
\label{fig:gbl}
\end{figure}

In Fig.\ref{fig:gbl}, we also show how $n_{s}$, $r$ and $dn_{s}/\ln k$ depend on $g_{B-L}(\Lambda_{\text{MPP}})$ when \\$\lambda_{\Psi}^{\text{eff}}(\Lambda_{\text{MPP}})=10^{-10}$. Here, we change $g_{B-L}(\Lambda_{\text{MPP}})$ within the region such that the electroweak symmetry breaking occurs at ${\cal{O}}(100)$GeV. Furthermore, $\Psi$ and $\xi$ are chosen so that they explain both of the observed value of  $A_{s}$ and $N=50$ when $g_{B-L}(\Lambda_{\text{MPP}})=0.0020$. One can see that $n_{s}$ and $r$ hardly depend on $g_{B-L}(\Lambda_{\text{MPP}})$ and that the change of $dn_{s}/\ln k$ is at most ${\cal{O}}(0.0001)$. As a result, under the situation where the minimum of the Higgs potential vanishes at $\Lambda_{\text{MPP}}$ and the electroweak symmetry breaking occurs at ${\cal{O}}(100)$GeV, the gauged B-L model uniquely predicts the cosmological observables. This is also one of the benefits of the (slightly broken) MPP. 

\section{Summary}\label{sec:sum}
In this paper, we have considered the MPP and the inflation of the gauged B-L extension of the SM. We have found that the scalar couplings and their beta functions can simultaneously become zero at $\Lambda_{\text{MPP}}=10^{17}$GeV and that the parameters of the model can be uniquely fixed by these conditions. However, from the point of view that the electroweak symmetry breaking should be realized by the radiatively broken B-L symmetry, it is necessary to break the MPP: we need $\lambda_{\Psi}^{\text{eff}}(\Lambda_{\text{MPP}})>0$ and $\beta_{\lambda_{\Psi}^{\text{eff}}}(\Lambda_{\text{MPP}})>0$. In subsection \ref{sub:breaking}, we found that the small values of $\lambda_{\Psi}^{\text{eff}}(\Lambda_{\text{MPP}})$ are compatible with the electroweak symmetry breaking at ${\cal{O}}(100)$GeV. In particular, we have found that the mass of the B-L gauge boson can be predicted to be 
\be M_{B-L}=2\sqrt{2}\times\sqrt{\frac{\lambda(v_{h})}{0.10}}\times v_{h}\e
from the MPP of the Higgs potential and $\kappa$. This is one of the remarkable predictions of the MPP. In Section \ref{sec:inf}, we have studied the inflation where the SM singlet scalar $\Psi$ plays a roll of the inflaton. We have calculated the cosmological observables based on the assumptions that the minimum of the Higgs potential vanishes at $\Lambda_{\text{MPP}}=10^{17}$GeV and the electroweak symmetry breaking occurs at ${\cal{O}}(100)$GeV. The results in this paper are consistent with the observations by Planck and BICEP2. Among them, the predicted values of the running of the spectral index $dn_{s}/\ln k$ are small compared with the observed values ${\cal{O}}(-0.01)$. It might be interesting to consider whether we can improve this situation. One of the such possibilities is to include a higher dimensional operator \cite{Hamada:2014wna}. In conclusion, the gauged B-L extension of the SM is a phenomenologically very interesting model in that it can explain both of the cosmological observations and the electroweak symmetry breaking at ${\cal{O}}(100)$GeV by breaking the MPP.

\section*{Acknowledgement} 
We thank Hikaru Kawai and Yuta Hamada for valuable discussions.

\appendix 
\def\thesection{Appendix \Alph{section}}
\section{Two-Loop Renormalization\\ \h{10cm} Group Equations}\label{app:beta}
The two-loop RGEs of the gauged B-L model are as follows\footnote{Our calculations are based on \cite{Machacek:1983tz,Machacek:1983fi,Machacek:1984zw,Fonseca:2013bua}. Especially, the two-loop results with an arbitrary number of Abelian groups are presented in \cite{Fonseca:2013bua}.}:
\begin{align}
\frac{d\Gamma_{H}}{dt}&=\frac{1}{(4\pi)^{2}}\left(\frac{9}{4}g_{2}^{2}+\frac{3}{4}g_{Y}^{2}+\frac{3}{4}g_{\text{mix}}^{2}-3y_{t}^{2}-3y_{\nu}^{2}\right),
\\
\frac{d\Gamma_{\Psi}}{dt}&=\frac{1}{(4\pi)^{2}}\left(12g_{B-L}^{2}-\frac{3}{2}Y_{R}^{2}\right),
\\
\nonumber\\
\frac{dg_{Y}}{dt}&=\frac{1}{(4\pi)^{2}}\frac{41}{6}g_{Y}^{3}+\frac{g_{Y}^{3}}{(4\pi)^{4}}\Biggl(\frac{199}{18}
   g_{Y}^2+\frac{9}{2} g_2^2+\frac{44}{3}
   g_3^2 +\frac{92}{9}
   g_{B-L}^2+\frac{199}{18} g_{\text{mix}}^2\nonumber
   \\
  &\h{9cm}+\frac{164}{9} g_{\text{mix}}
   g_{B-L}
   -\frac{17}{6}  y_t^2-\frac{3}{2}
    y_{\nu }^2\Biggl),\end{align}
\begin{align}
\frac{dg_{\text{mix}}}{dt}&=\frac{1}{(4\pi)^{2}}\left(\frac{41}{6}g_{\text{mix}}\left(g_{\text{mix}}^{2}+2g_{Y}^{2}\right)+\frac{32}{3}g_{B-L}\left(g_{\text{mix}}^{2}+g_{Y}^{2}\right)+12g_{\text{mix}}g_{B-L}^{2}\right)\nonumber
\\
&+\frac{1}{(4\pi)^{4}}\Biggl\{g_{\text{mix}}^3\left(\frac{199}{18}
   g_{\text{mix}}^2+\frac{328}{9}
   g_{\text{mix}} g_{B-L}+\frac{9}{2} g_2^2
   +\frac{44}{3} g_3^2
   +\frac{184}{3}
    g_{B-L}^2+\frac{199}{6}
   g_{Y}^2\right)\nonumber
   \\
   &\quad+g_{\text{mix}}^2\Biggl(\frac{656}{9} 
   g_{Y}^2 g_{B-L}+\frac{448}{9}
    g_{B-L}^3+\frac{32}{3} g_3^2  
   g_{B-L}+12 g_2^2
   g_{B-L}\Biggl)\nonumber
   \\
   &\quad+g_{\text{mix}}\Biggl(\frac{644}{9}
   g_{Y}^2 g_{B-L}^2+\frac{800}{9}
   g_{B-L}^4+12 g_2^2
   g_{B-L}^2+\frac{199}{9}g_{Y}^4+9
   g_2^2  g_{Y}^2+\frac{88}{3}
   g_3^2  g_{Y}^2+\frac{32}{3} g_3^2
   g_{B-L}^2\Biggl)\nonumber
   \\
   &\quad+\frac{164}{9} g_{Y}^4 g_{B-L}+\frac{224}{9}
   g_{Y}^2 g_{B-L}^3+12 g_2^2 g_{Y}^2
   g_{B-L}+\frac{32}{3} g_3^2 g_{Y}^2
   g_{B-L}\nonumber
   \\
   &\quad-y_t^2\left(\frac{10}{3} g_{Y}^2
    g_{B-L}+\frac{17}{6} g_{\text{mix}}^3
   +\frac{10}{3} g_{\text{mix}}^2
   g_{B-L}+\frac{4}{3} g_{\text{mix}}
   g_{B-L}^2+\frac{17}{3} g_{\text{mix}} g_{Y}^2
   \right)\nonumber
   \\
   &\quad-y_{\nu }^2\left(6 g_{Y}^2 
   g_{B-L}+\frac{3}{2} g_{\text{mix}}^3+3 g_{\text{mix}} g_{Y}^2+6 g_{\text{mix}}^2 
   g_{B-L}+12 g_{\text{mix}} 
   g_{B-L}^2\right)-3Y_R^2g_{\text{mix}}
   g_{B-L}^2 \Biggl\},
\end{align}

\begin{align}
\frac{dg_{B-L}}{dt}&=\frac{g_{B-L}}{(4\pi)^{2}}\left(12g_{B-L}^{2}+\frac{32}{3}g_{B-L}g_{\text{mix}}+\frac{41}{6}g_{\text{mix}}^{2}\right)\nonumber
\\
&+\frac{g_{B-L}}{(4\pi)^{4}}\Biggl\{g_{B-L}^2\left(\frac{800}{9} g_{B-L}^2+\frac{92}{9} g_y^2+\frac{184}{3}
   g_{\text{mix}}^2
   +12 g_2^2
   +\frac{32}{3} g_3^2+\frac{448}{9}
   g_{\text{mix}} g_{B-L}\right)\nonumber
   \\
   & \quad+g_{B-L}\left(\frac{164}{9} g_{\text{mix}}
   g_y^2 +\frac{328}{9}
   g_{\text{mix}}^3 +12 g_2^2
   g_{\text{mix}} +\frac{32}{3} g_3^2
   g_{\text{mix}}\right)\nonumber
   \\
   &\quad+\frac{199}{18}
   g_{\text{mix}}^2 g_y^2+\frac{199}{18}
   g_{\text{mix}}^4 +\frac{9}{2} g_2^2
   g_{\text{mix}}^2+\frac{44}{3} g_3^2
   g_{\text{mix}}^2\nonumber
   \\
   &\quad-y_t^2\left(\frac{4}{3}g_{B-L}^2+\frac{10}{3} g_{\text{mix}}
   g_{B-L}+\frac{17}{6} g_{\text{mix}}^2 \right)
   -y_{\nu }^2\left(6 g_{\text{mix}} 
   g_{B-L}-\frac{3}{2} g_{\text{mix}}^2 -12
   g_{B-L}^2\right)-3Y_R^2
   g_{B-L}^2\Biggl\},
\end{align}

\be
\frac{dg_{2}}{dt}=-\frac{1}{(4\pi)^{2}}\frac{19}{6}g_{2}^{3}+\frac{g_{2}^{3}}{(4\pi)^{4}}\left(\frac{3}{2}  g_{Y}^2+\frac{35}{6}
   g_2^2+12 g_3^2 +4
   g_{B-L}^2+\frac{3}{2} 
   g_{\text{mix}}^2+4  g_{\text{mix}} g_{B-L}-\frac{3}{2}
   y_t^2-\frac{3}{2}  y_{\nu
   }^2\right),
\e
\be
\frac{dg_{3}}{dt}=-\frac{7}{(4\pi)^{2}}g_{3}^{3}+\frac{g_{3}^{3}}{(4\pi)^{4}}\left(\frac{11}{6}
    g_{Y}^2+\frac{9}{2} g_2^2-26 g_3^2+\frac{4}{3}
    g_{B-L}^2+\frac{11}{6} 
   g_{\text{mix}}^2+\frac{4}{3}  g_{\text{mix}} g_{B-L}-2  y_t^2
   \right),
\e
\begin{align}
\frac{dy_{t}}{dt}&=\frac{y_{t}}{(4\pi)^{2}}\left(\frac{9}{2}y_{t}^{2}+3y_{\nu}^{2}-8g_{3}^{2}-\frac{9}{4}g_{2}^{2}-\frac{17}{12}g_{Y}^{2}-\frac{17}{12}g_{\text{mix}}^{2}-\frac{2}{3}g_{B-L}^{2}-\frac{5}{3}g_{B-L}g_{\text{mix}}\right)\nonumber
\\
&+\frac{y_{t}}{(4\pi)^{4}}\Biggl\{-12
   y_t^4-\frac{27}{4} y_{\nu }^4-\frac{27}{4} y_t^2
   y_{\nu }^2-\frac{9}{4} Y_R^2 y_{\nu
   }^2+6 \lambda ^2+\frac{1}{2} \kappa ^2-12 \lambda 
   y_t^2\nonumber
   \\
   &\quad+y_t^2\left(36 g_3^2+\frac{225}{16} g_2^2
   +\frac{131}{16} g_{Y}^2 +3
   g_{B-L}^2+\frac{131}{16} g_{\text{mix}}^2
   +\frac{25}{4} g_{\text{mix}}
   g_{B-L}\right)\nonumber
   \\
   &\quad+y_{\nu
   }^2\left(\frac{45}{8}g_{2}^{2} +\frac{15}{8} g_{Y}^2 +15g_{B-L}^2+\frac{15}{8} g_{\text{mix}}^2
   +\frac{15}{2} g_{\text{mix}}
   g_{B-L}\right)\nonumber
   \\
   &\quad+\frac{502}{27} g_{\text{mix}}^3
   g_{B-L}+\frac{1085}{36} g_{\text{mix}}^2
   g_{B-L}^2+\frac{502}{27} g_{\text{mix}} g_{Y}^2
   g_{B-L}+\frac{665}{27} g_{\text{mix}}
   g_{B-L}^3+\frac{9}{4} g_2^2 g_{\text{mix}}
   g_{B-L}\nonumber
   \\
   &\quad-\frac{20}{9} g_3^2 g_{\text{mix}}
   g_{B-L}+\frac{203}{27}
   g_{B-L}^4+\frac{3}{4} g_2^2
   g_{B-L}^2-\frac{8}{9} g_3^2
   g_{B-L}^2+\frac{91}{12} g_{Y}^2
   g_{B-L}^2+\frac{1187}{216} g_{\text{mix}}^4
   -\frac{3}{4} g_2^2 g_{\text{mix}}^2
   \nonumber
   \\
   &\quad+\frac{19}{9} g_3^2 g_{\text{mix}}^2
   +\frac{1187}{108} g_{\text{mix}}^2 g_{Y}^2
  -\frac{23}{4}
   g_2^4-108 g_3^4+\frac{1187}{216}
   g_{Y}^4+9 g_2^2 g_3^2-\frac{3}{4}
   g_2^2 g_{Y}^2+\frac{19}{9} g_3^2
   g_{Y}^2\Biggl\},
\end{align}
\begin{align}
\frac{dy_{\nu}}{dt}&=\frac{y_{\nu}}{(4\pi)^{2}}\left(-3 g_{\text{mix}} g_{B-L}-6 g_{B-L}^2-\frac{3}{4}
   g_{\text{mix}}^2-\frac{3}{4} g_{Y}^2-\frac{9}{4}
   g_2^2+\frac{1}{4} Y_R^2+3 y_t^2+\frac{9}{2}
   y_{\nu }^2\right)\nonumber
   \\
&+\frac{y_{\nu}}{(4\pi)^{4}}\Biggl\{-12 y_{\nu }^4-\frac{27}{4}
   y_t^4-\frac{5}{4} Y_R^4-y_{\nu }^2\left(\frac{27}{4} y_t^2+\frac{21}{8}
   Y_R^2  \right) +6 \lambda ^2+\frac{1}{2} \kappa ^2-12 \lambda  y_{\nu }^2-\kappa  Y_R^2\nonumber
   \\
   &\quad+y_{\nu}^{2}\left(\frac{225}{16} g_2^2+\frac{123}{16} g_{Y}^2 +27 g_{B-L}^2+\frac{123}{16} g_{\text{mix}}^2
 +\frac{69}{4} g_{\text{mix}} g_{B-L}\right)\nonumber
   \\
   &\quad+y_{t}^{2}\left(20
   g_3^2+\frac{45}{8} g_2^2+\frac{85}{24} g_{Y}^2
   +\frac{5}{3}
   g_{B-L}^2+\frac{85}{24} g_{\text{mix}}^2
   +\frac{25}{6} g_{\text{mix}} g_{B-L}\right)+Y_{R}^{2}\left(22
   g_{B-L}^2+3 g_{\text{mix}} 
   g_{B-L}\right)\nonumber
   \\
   &\quad+21 g_{\text{mix}}^3
   g_{B-L}+\frac{799}{12} g_{\text{mix}}^2
   g_{B-L}^2+21 g_{\text{mix}} g_{Y}^2
   g_{B-L}+\frac{253}{3} g_{\text{mix}}
   g_{B-L}^3+\frac{9}{4} g_2^2 g_{\text{mix}}
    g_{B-L}+65 g_{B-L}^4\nonumber
    \\
    &\quad+\frac{27}{4} g_2^2
   g_{B-L}^2+\frac{187}{12} g_{Y}^2
   g_{B-L}^2+\frac{35}{24} g_{\text{mix}}^4 -\frac{9}{4} g_2^2 g_{\text{mix}}^2 +\frac{35}{12} g_{\text{mix}}^2 g_{Y}^2-\frac{23}{4} g_2^4+\frac{35}{24}
   g_{Y}^4 -\frac{9}{4} g_2^2 g_{Y}^2
  \Biggl\},
\end{align}
\begin{align}
\frac{dY_{R}}{dt}&=\frac{Y_{R}}{(4\pi)^{2}}\left(\frac{5}{2}Y_{R}^{2}+2y_{\nu}^{2}-6g_{B-L}^{2}\right)\nonumber
\\
&+\frac{Y_{R}}{(4\pi)^{4}}\Biggl\{-5Y_R^4-\frac{19}{2} Y_R^2 y_{\nu
   }^2-9
   y_t^2 y_{\nu }^2-\frac{27}{2} y_{\nu }^4+4 \lambda
   _{\Psi }^2+\kappa ^2-8 \kappa  y_{\nu
   }^2
   -8 \lambda _{\Psi }Y_R^2\nonumber
   \\
   &\quad+y_{\nu }^2\left(\frac{51}{4} g_2^2 +\frac{17}{4} g_{Y}^2+16
   g_{B-L}^2+\frac{17}{4} g_{\text{mix}}^2
   +11 g_{\text{mix}}
   g_{B-L}\right)+\frac{103}{2} Y_R^2 g_{B-L}^2\nonumber
   \\
   &\quad-127
    g_{B-L}^4-\frac{35}{6} g_{\text{mix}}^2
   g_{B-L}^2-\frac{32}{3} g_{\text{mix}}
   g_{B-L}^3\Biggl\},
\end{align}
\begin{align}\frac{d\lambda_{\Psi}}{dt}
&=\frac{1}{(4\pi)^{2}}\left(\lambda _{\Psi
   }\left(20 \lambda _{\Psi
   }-48
   g_{B-L}^2+6 Y_R^2\right)+2 \kappa ^2+96 g_{B-L}^4-3 Y_R^4\right)\nonumber
   \\
&+\frac{1}{(4\pi)^{4}}\Biggl\{-240 \lambda _{\Psi }^3-20 \kappa ^2 \lambda _{\Psi }-8 \kappa
   ^3+\lambda
   _{\Psi }\left(\frac{1280}{3} g_{\text{mix}}  g_{B-L}^3+\frac{844}{3}
   g_{\text{mix}}^2
   g_{B-L}^2+2112 g_{B-L}^4\right)\nonumber
   \\
   &\quad+448
   \lambda _{\Psi }^2 g_{B-L}^2-7168 g_{B-L}^6-\frac{8192}{3} g_{\text{mix}}
   g_{B-L}^5-\frac{5344}{3} g_{\text{mix}}^2
   g_{B-L}^4+12 Y_R^4
   g_{B-L}^2+288 Y_R^2 g_{B-L}^4\nonumber
   \\
   &\quad+30 \lambda _{\Psi }Y_R^2
   g_{B-L}^2-60
   Y_R^2 \lambda _{\Psi }^2+ \lambda _{\Psi
   } \left( 3Y_R^4-18 Y_R^2y_{\nu }^2
   \right)+12 Y_R^4 y_{\nu }^2+12 Y_R^6-12 \kappa^{2}\left(
   y_t^2+y_{\nu
   }^2\right)\nonumber
   \\
   &\quad+\kappa
   ^2\left(12 g_2^2 +4 g_{Y}^2+4
   g_{\text{mix}}^2\right)+40 \kappa  g_{\text{mix}}^2
   g_{B-L}^2\Biggl\},
\end{align}
\begin{align}
\frac{d\lambda}{dt}&=\frac{1}{16\pi^{2}}\Biggl(\lambda\left(24
   \lambda-9
   g_{2}^{2}-3
   g_{\text{mix}}^{2}-3  g_{Y}^{2}+12
   y_{t}^{2}+12 y_{\nu }^{2}\right)+\frac{3}{4} \left(g_{Y}^{2}+g_{\text{mix}}^{2} \right)g_{2}^{2}+\frac{3}{4} g_{\text{mix}}^{2}
   g_{Y}^{2}\nonumber\\
   &\quad\quad\quad\quad+\frac{9}{8} g_{2}^{4}+\frac{3}{8} g_{\text{mix}}^{4}+\frac{3}{8}
   g_{Y}^{4}+\kappa^{2}-6 y_{t}^{4}-6y_{\nu}^{4}\Biggl)\nonumber
\\  
  &+\frac{1}{(4\pi)^{4}}\Biggl\{-4 \kappa ^3-10 \kappa ^2 \lambda-312 \lambda ^3+36 \lambda ^2
   \left(g_{Y}^2+g_{\text{mix}}^2+3
   g_2^2\right)+\lambda\Biggl(\frac{629}{24}
   g_{Y}^4+\frac{629}{24}
   g_{\text{mix}}^4+\frac{39}{4}  g_{\text{mix}}^2
   g_2^2\nonumber 
   \\
   &\quad+\frac{39}{4}  g_{Y}^2 
   g_2^2-\frac{73}{8} 
   g_2^4+\frac{80}{3}
   g_{B-L} g_{\text{mix}}^3+34 
    g_{B-L}^2 g_{\text{mix}}^2+\frac{629}{12}  g_{\text{mix}}^2
   g_{Y}^2+\frac{80}{3}   g_{B-L}
   g_{\text{mix}} g_{Y}^2\Biggl)\nonumber
   \\
   &\quad+\frac{305}{16} g_2^6-\frac{289}{48}
   g_{\text{mix}}^2 g_2^4-\frac{289}{48} g_{Y}^2
   g_2^4-\frac{559}{48} g_{\text{mix}}^4
   g_2^2-\frac{559}{48} g_{Y}^4 g_2^2-\frac{32}{3}
   g_{B-L} g_{\text{mix}}^3 g_2^2\nonumber
   \\
   & \quad-13 g_{B-L}^2
   g_{\text{mix}}^4-\frac{379}{16} g_{\text{mix}}^2
   g_{Y}^4-\frac{32}{3} g_{B-L} g_{\text{mix}}
   g_{Y}^4-\frac{32}{3} g_{B-L}
   g_{\text{mix}}^5-13 g_{B-L}^2 g_{\text{mix}}^2 
   g_2^2-\frac{559}{24} g_{\text{mix}}^2 g_{Y}^2
   g_2^2\nonumber
   \\
  &\quad-\frac{379}{48}
   g_{\text{mix}}^6-\frac{379}{48} g_{Y}^6-\frac{32}{3} g_{B-L} g_{\text{mix}}
   g_{Y}^2 g_2^2-\frac{379}{16}
   g_{\text{mix}}^4 g_{Y}^2-\frac{64}{3} g_{B-L}
   g_{\text{mix}}^3 g_{Y}^2-13 g_{B-L}^2 g_{\text{mix}}^2
   g_{Y}^2\nonumber
   \\
   &\quad-y_t^4\left(\frac{8}{3} g_{Y}^2 +32 g_3^2 +\frac{8}{3} 
   g_{B-L}^2 +\frac{8}{3} g_{\text{mix}}^2
  +\frac{20}{3} g_{B-L} g_{\text{mix}}
    \right)-y_{\nu }^4\left(24 g_{B-L}^2 +12 g_{B-L} g_{\text{mix}} \right)\nonumber 
   \\
  &\quad+y_t^2\Biggl(\frac{21}{2} g_{\text{mix}}^2 
   g_2^2+\frac{21}{2} g_{Y}^2 
   g_2^2+6
   g_{B-L} g_{\text{mix}} 
   g_2^2-\frac{9}{4} g_2^4-\frac{19}{4}
   g_{\text{mix}}^4-\frac{19}{4} g_{Y}^4
    -\frac{19}{2}
   g_{\text{mix}}^2 g_{Y}^2-4 g_{B-L}^2
   g_{\text{mix}}^2\nonumber
   \\
  &\h{10cm} -10 g_{B-L} g_{\text{mix}}^3
   -10 g_{B-L}
   g_{\text{mix}} g_{Y}^2 \Biggl)\nonumber
   \\
   &\quad-y_{\nu }^2\Biggl(\frac{3}{2} g_{\text{mix}}^2  
   g_2^2+\frac{3}{2} g_{Y}^2 
   g_2^2+18 g_{B-L} g_{\text{mix}} 
   g_2^2+\frac{9}{4}
    g_2^4+\frac{3}{4}
   g_{\text{mix}}^4 +\frac{3}{4} g_{Y}^4
   +18 g_{B-L} g_{\text{mix}}^3 +36 g_{B-L}^2
   g_{\text{mix}}^2\nonumber
   \\
  &\h{10cm} +\frac{3}{2}
   g_{\text{mix}}^2 g_{Y}^2+18 g_{B-L} g_{\text{mix}} g_{Y}^2 \Biggl)\nonumber
   \\
   &\quad+\lambda y_{t}^{2}\left(\frac{45}{2}  g_2^2+80 
   g_3^2  +\frac{85}{6}
   g_{Y}^2+\frac{85}{6}
   g_{\text{mix}}^2 +\frac{20}{3} 
   g_{B-L}^2+\frac{50}{3}
    g_{B-L} g_{\text{mix}} \right)\nonumber
   \\
  & \quad+\lambda y_{\nu}^{2}\left(\frac{45}{2} 
   g_2^2+\frac{15}{2} g_{Y}^2  +60  
   g_{B-L}^2 +\frac{15}{2} 
    g_{\text{mix}}^2+30 g_{B-L} g_{\text{mix}} \right)
   \nonumber
   \\
   &\quad-144 \lambda ^2 \left(y_t^2+y_{\nu }^2\right)-3 \lambda\left( 
   y_t^4+ 
   y_{\nu }^4+3 y_{\nu }^2
   Y_R^2 \right)+30 
   y_t^6+30 y_{\nu }^6+12 y_{\nu }^4 Y_R^2-3\kappa
   ^2 Y_R^2\nonumber
   \\
   &\quad+32 \kappa ^2
   g_{B-L}^2+20
   \kappa  g_{B-L}^2 g_{\text{mix}}^2
   \Biggl\},\label{eq:betalam}
\end{align}
\begin{align}
\frac{d\kappa}{dt}&=\frac{1}{(4\pi)^{2}}\Biggl\{\kappa\left(4
   \kappa+12\lambda+8
   \lambda _{\Psi } -\frac{3}{2}g_y^2-24
   g_{B-L}^2-\frac{3}{2}
   g_{\text{mix}}^2-\frac{9}{2} g_2^2
  +3 Y_R^2+6y_t^2 +6  y_{\nu }^2\right)\nonumber
  \\
  &\h{11cm}+12 g_{\text{mix}}^2 g_{B-L}^2-12 Y_R^2 y_{\nu
   }^2\Biggl\}\nonumber
   \\
   &+\frac{1}{(4\pi)^{4}}\Biggl\{\kappa\Biggl(-11 \kappa 
   ^2-40
   \lambda _{\Psi }^2-48 \kappa \lambda _{\Psi }-60 \lambda ^2-72 \kappa \lambda-(y_t^2+y_{\nu}^2)\left(12
   \kappa +72
   \lambda \right)-Y_R^2\left(6 \kappa
   +24\lambda _{\Psi }\right)\nonumber
   \\
   &\quad-\frac{27}{2} y_t^4-\frac{27}{2}
   y_{\nu }^4-\frac{9}{2}  Y_R^4+\frac{21}{2}
    y_{\nu }^2 Y_R^2+g_2^2\left(3 \kappa
   +72 \lambda\right)+ g_{Y}^2\left(\kappa
  +24  \lambda 
   \right)+g_{B-L}^2\left(16 \kappa
    +256
    \lambda _{\Psi }\right)\nonumber 
 \\
   &\quad+g_{\text{mix}}^2\left(\kappa
   +24  \lambda
    \right)+y_t^2\left(\frac{45}{4} 
   g_2^2+40   g_3^2
   +\frac{10}{3} g_{B-L}^2
   +\frac{85}{12} g_{\text{mix}}^2
   +\frac{85}{12}  g_{Y}^2 +\frac{25}{3}
   g_{B-L} g_{\text{mix}}\right)\nonumber
   \\
   &\quad+y_{\nu }^2\left(\frac{45}{4} 
   g_2^2+30
    g_{B-L}^2+\frac{15}{4}
    g_{\text{mix}}^2 +\frac{15}{4}
   g_{Y}^2 +15 g_{B-L} g_{\text{mix}}
   \right)+15 Y_R^2g_{B-L}^2 \nonumber 
   \\
   &\quad-\frac{145}{16} g_2^4+\frac{557}{48} 
   g_{Y}^4+672
    g_{B-L}^4+\frac{557}{48}
   g_{\text{mix}}^4+\frac{15}{8}  g_{\text{mix}}^2
   g_2^2+\frac{15}{8}  g_{Y}^2
   g_2^2+\frac{40}{3} g_{B-L}
   g_{\text{mix}}^3\nonumber
   \\
   &\quad+\frac{497}{3}
   g_{B-L}^2 g_{\text{mix}}^2+\frac{557}{24}
   g_{\text{mix}}^2 g_{Y}^2+\frac{40}{3}  g_{B-L}
   g_{\text{mix}} g_{Y}^2+\frac{640}{3} g_{B-L}^3
   g_{\text{mix}}\Biggl)\nonumber
   \\
   &\quad+y_{\nu }^2 Y_R^2\left(30 Y_R^2+66 y_{\nu }^2-84 
   g_{B-L}^2 -\frac{3}{2}
   g_{\text{mix}}^2 -\frac{3}{2}
   g_{Y}^2-24 g_{B-L}
   g_{\text{mix}} +\frac{9}{2}
   g_2^2\right)\nonumber
   \\
   &\quad-y_t^2\left(64 g_{B-L}^4 +76 g_{B-L}^2 g_{\text{mix}}^2
   +160 g_{B-L}^3
   g_{\text{mix}}\right)-y_{\nu }^2\left(576 g_{B-L}^4
   +12 g_{B-L}^2
   g_{\text{mix}}^2 +288 g_{B-L}^3 g_{\text{mix}}
   \right)\nonumber
   \\
   &\quad-18
   g_{B-L}^2 g_{\text{mix}}^2 Y_R^2+80\lambda _{\Psi }
   g_{B-L}^2 g_{\text{mix}}^2 +120 \lambda 
   g_{B-L}^2 g_{\text{mix}}^2\nonumber
   \\
   &\quad-45 g_{B-L}^2 g_{\text{mix}}^2
   g_2^2-\frac{713}{3} g_{B-L}^2
   g_{\text{mix}}^4-\frac{1024}{3} g_{B-L}^3
   g_{\text{mix}}^3-656 g_{B-L}^4
   g_{\text{mix}}^2-\frac{713}{3} g_{B-L}^2
   g_{\text{mix}}^2 g_{Y}^2\nonumber
   \\
  &\h{12cm}-\frac{512}{3} g_{B-L}^3 g_{\text{mix}}
   g_{Y}^2\Biggl\}.\label{eq:betak}
\end{align}

\end{document}